\documentclass[onecolumn, journal,11pt]{IEEEtran}

\usepackage{xspace}
\usepackage{mathrsfs}
\usepackage{amsopn}

\def\Var{\mathop{\rm Var}\nolimits}%
\newcommand{\Ac}{\mathcal{A}}
\newcommand{\Bc}{\mathcal{B}}
\newcommand{\Cc}{\mathcal{C}}

\newcommand{\Ec}{\mathcal{E}}

\newcommand{\Rc}{\mathcal{R}}

\newcommand{\Uc}{\mathcal{U}}
\newcommand{\Vc}{\mathcal{V}}

\newcommand{\Xc}{\mathcal{X}}
\newcommand{\Yc}{\mathcal{Y}}
\newcommand{\Zc}{\mathcal{Z}}

\newcommand{\aep}{{\mathcal{T}_{\epsilon}^{(n)}}}
\newcommand{\aepvar}{{\mathcal{T}_{\epsilon'}^{(n)}}}

\newcommand{\Lh}{{\hat{L}}}

\newcommand{\Xh}{{\hat{X}}}

\newcommand{\lh}{{\hat{l}}}
\newcommand{\mh}{{\hat{m}}}

\newcommand{\xh}{{\hat{x}}}

\newcommand{\Rt}{{\tilde{R}}}

\newcommand{\Wt}{{\tilde{W}}}
\newcommand{\Xt}{{\tilde{X}}}
\newcommand{\Yt}{{\tilde{Y}}}
\newcommand{\Zt}{{\tilde{Z}}}

\newcommand{\wt}{{\tilde{w}}}
\newcommand{\xt}{{\tilde{x}}}
\newcommand{\yt}{{\tilde{y}}}
\newcommand{\zt}{{\tilde{z}}}

\def\d{\delta}
\def\e{\epsilon}

\DeclareMathOperator\E{E}
\let\P\relax
\DeclareMathOperator\P{P}

\newcommand{\Bern}{\mathrm{Bern}}

\newcommand{\U}{\mathcal{U}}

\def\textiid{i.i.d.\@\xspace}
\newcommand\iid{\ifmmode\text{ i.i.d. } \else \textiid \fi}

\usepackage{cite, amsmath, amssymb, epsfig,psfrag,color,subfig, fullpage}

\newtheorem{remark}{Remark}[section]
\newtheorem{lemma}{Lemma}

\newtheorem{proposition}{Proposition}
\newtheorem{corollary}{Corollary}

\begin{document}
\title{On Secure Source Coding with Side Information at the Encoder}

\author{Yeow-Khiang Chia\IEEEauthorrefmark{1} and Kittipong Kittichokechai\IEEEauthorrefmark{2}
\thanks{To be presented in part at IEEE International Symposium for Information Theory 2013}
\thanks{\IEEEauthorrefmark{1} Yeow-Khiang Chia is with Institute for Infocomm Research, Singapore. Email: yeowkhiang@gmail.com}  \thanks{\IEEEauthorrefmark{2} Kittipong Kittichokechai is with ACCESS Linnaeus Center, KTH Royal Institute of Technology, Sweden. Email: kki@kth.se}%
}
\maketitle
\vspace{-25pt}
\begin{abstract}
We consider a secure source coding problem with side information (S.I.) at the decoder and the eavesdropper. The encoder has a source that it wishes to describe with limited distortion through a rate limited link to a legitimate decoder. The message sent is also observed by the eavesdropper. The encoder aims to minimize both the distortion incurred by the legitimate decoder; and the information leakage rate at the eavesdropper. When the encoder has access to the uncoded S.I. at the decoder, we characterize the rate-distortion-information leakage rate (R.D.I.) region under a Markov chain assumption and when S.I. at the encoder does not improve the rate-distortion region as compared to the case when S.I. is absent. When the decoder also has access to the eavesdropper's S.I., we characterize the R.D.I. region without the Markov Chain condition. We then consider a related setting where the encoder and decoder obtain coded S.I. through a rate limited helper, and characterize the R.D.I. region for several special cases, including special cases under logarithmic loss distortion and for special cases of the Quadratic Gaussian setting. Finally, we consider the amplification measures of list or entropy constraint at the decoder, and show that the R.D.I. regions for the settings considered in this paper under these amplification measures coincide with R.D.I. regions under per symbol logarithmic loss distortion constraint at the decoder.   
\end{abstract}
\section{Introduction}
\begin{figure}
\begin{center}
\psfrag{Encoder}[c]{Enc.}
\psfrag{Eav}[c]{Eaves.}
\psfrag{Dec}[c]{Dec.}
\psfrag{Enc}[c]{Enc}
\psfrag{X}[c]{$X^n$}
\psfrag{r}[c]{$R$}
\psfrag{Z}[c]{$Z^n$}
\psfrag{Y}[c]{$Y^n$}
\psfrag{rh}[c]{$R_h$}
\psfrag{Dis}[c]{$\Xh^n$}
\psfrag{W}[c]{$W^n$}
\psfrag{Helper}[l]{Helper}
\psfrag{IL}[c]{$\frac{I(X^n; M, Z^n)}{n}$}
\psfrag{IL2}[c]{$\frac{I(X^n; M, W^n)}{n}$}
{\includegraphics[width = 0.75\linewidth, height = 0.3\linewidth]{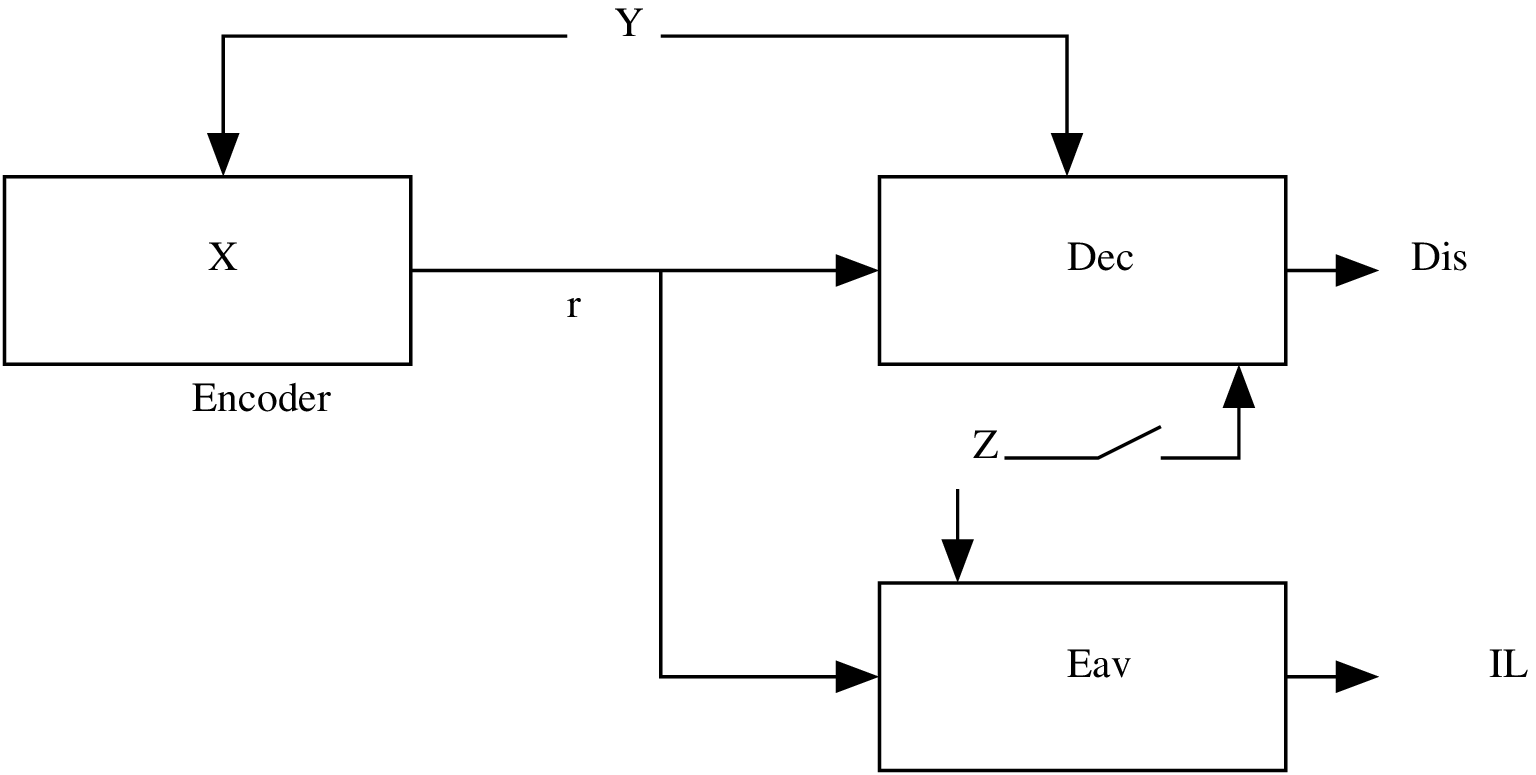}}
\caption{Uncoded S.I. at the encoder. When the switch is opened, this figure corresponds to the setting described in Section~\ref{sect:2A}. When the switch is closed, this figure corresponds to the setting described in Section~\ref{sect:2B}} \label{fig1}
\vspace{5pt}
{\includegraphics[width = 0.75\linewidth, height = 0.3\linewidth]{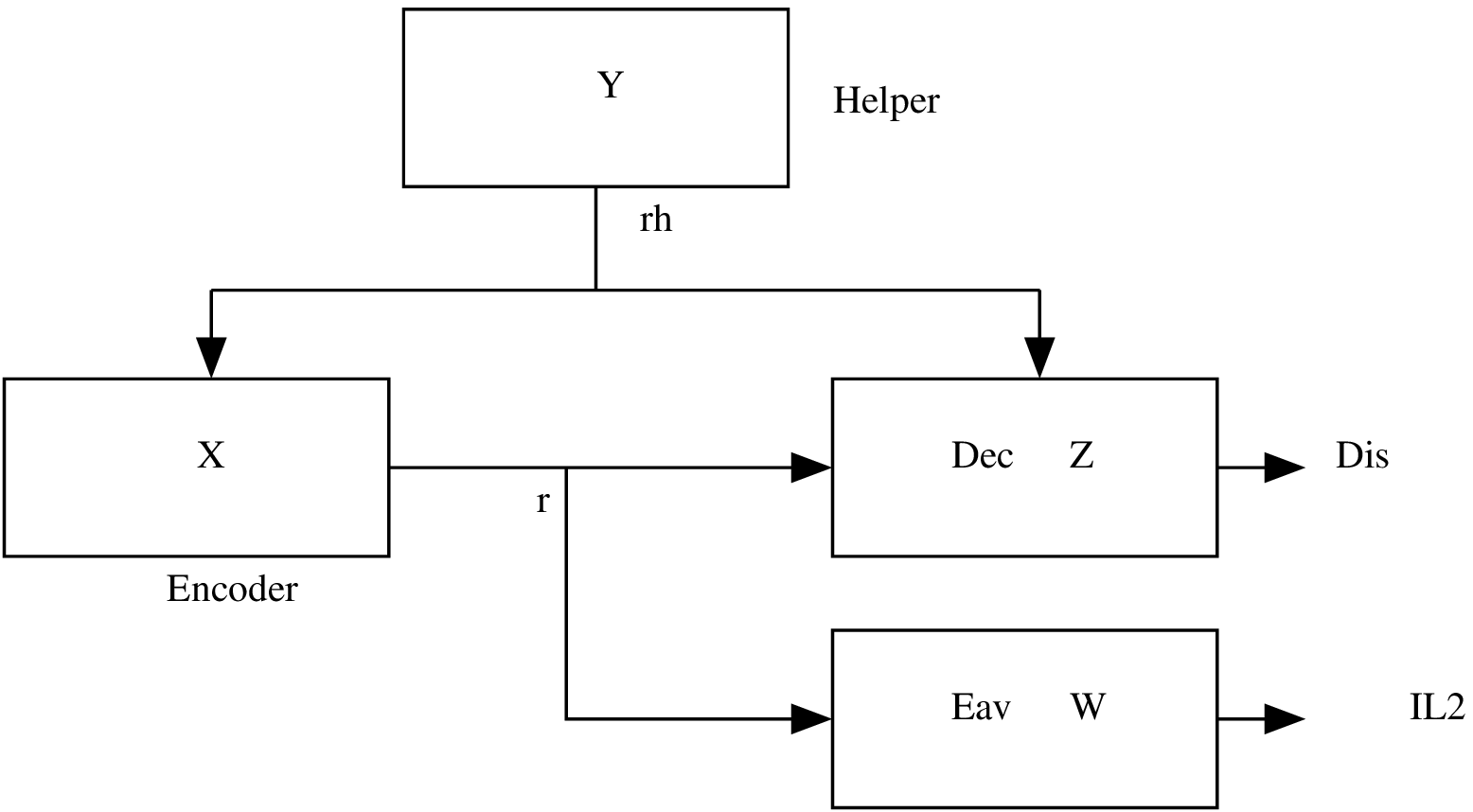}}
\caption{Coded S.I. at the encoder.} \label{fig2}
\end{center}
\end{figure}

Consider the secure lossy source coding problem with S.I. at the decoders in Figure~\ref{fig1}. The encoder has source $X^n$ that it wishes to describe lossily through a rate limited link to decoder 1 (legitimate decoder). The message sent is also observed by decoder 2, which is an eavesdropper in our setup. The encoder aims to minimize the distortion incurred by decoder 1 in reconstructing the source sequence, while at the same time, minimize the information leakage rate at the eavesdropper given its S.I. and the common message $M$: $I(X^n; M, Z^n)/n$.


The problem of source coding with security constraints has received attention in recent years,~\cite{Gunduz2008, Tandon2009, Villard2010, Tandon2012}, due to potential applications in areas such as privacy in sensor networks and databases. For example,~\cite{Shankar2012} approached the issue of privacy in databases from an information theoretic perspective, using the information leakage rate as a privacy measure. The use of the information leakage rate as a measure of privacy has also found applications in the area of smart grid, and in particular, privacy for smart meters. We refer interested readers to~\cite{TanGunduzPoor2013},~\cite{Rajaopalan2011},~\cite{Varodayan2011} and the references therein for work in this area. Among the literature on secure source coding, of particular relevance to this work are the papers~\cite{Villard2010} and~\cite{Tandon2012}. In~\cite{Villard2010}, the authors considered our setting when S.I. $Y^n$ is unavailable at the encoder and gave the full characterization of the rate-distortion-information leakage rate (R.D.I.) region for discrete memoryless sources and arbitrary distortion measures.~\cite{Tandon2012} considered both the case when S.I. $Y^n$ is available at the encoder and the case when S.I. $Y^n$ is unavailable at encoder. However, the authors were interested in the information leakage rate for the S.I., $I(Y^n; M, Z^n)/n$, instead of $I(X^n; M, Z^n)/n$. As we will discuss in the sequel, the differences give rise to a new role, that of generating secret key from common randomness~\cite{Ahlswede--Csiszar1993}, for the S.I. observed at the encoder and decoder.

A particular distortion measure that we will focus on in this paper is the \textit{logarithmic loss} (log-loss) distortion measure, first proposed in~\cite{Courtade2011}. Log-loss has the interesting property that S.I. at the encoder does not improve the rate-distortion region, with respect to the Wyner-Ziv setting~\cite{Wynerrd} where S.I. is absent at the encoder. This property will be key in establishing the results in this paper. Following~\cite{Coutrade2012a} and~\cite{Courtade2012c}, we will also extend our work to consider source amplification measures for our setting. We consider the amplification measures of list constraint~\cite{Kim--Sutivong--Cover2008}, and the block entropy constraint, $H(X^n|M, Y^n)/n$, at the decoder. Interestingly, we find, for our settings, that the R.D.I. region is the same regardless of whether one uses symbol by symbol log-loss or the above amplification measures.   

The rest of this paper is as follow. We first provide formal definitions in Section~\ref{sect:2}. Our main results are then given in the subsequent sections and summarized here:
\begin{itemize}
\item In Section~\ref{sect:3}, we consider our setting in Figure~\ref{fig1} when the eavesdropper's S.I. is not available at the legitimate decoder. General inner and outer bounds are given for this setup and the R.D.I. region is characterized when these conditions hold: (i) a Markov Chain $X-Y-Z$ between the source and the side informations; and (ii) S.I. at the encoder does not improve the rate distortion region. 
\item Section~\ref{sect:4} considers the setting where the eavesdropper's S.I. is available at the decoder (Figure~\ref{fig1} with the switch closed). We characterize the R.D.I. region when S.I. at the encoder does not improve the rate distortion region. 
\item Section~\ref{sect:5} considers the setting in Figure~\ref{fig2}, where the encoder and decoder obtain \textit{coded} S.I. sent by a helper via a rate-limited link. We present a general achievability scheme for this setting and show that the achievability scheme is optimal for some distortion measures when the source and S.I.s satisfy certain Markov Chain conditions. We also extend our analysis for this setting to the Quadratic Gaussian case, and characterize the R.D.I. regions for some special cases.
\item In Section~\ref{sect:6}, we consider the amplification measures listed in the previous paragraph, and show that the R.D.I. regions under the amplification measures are the same as that under log-loss for the settings considered in this paper.  
\end{itemize}
Finally, we conclude the paper in Section~\ref{sect:7}. 
\section{Definitions} \label{sect:2}
We will follow the notation in~\cite{El-Gamal--Kim2010}. Throughout this paper, source and side informations $(X^n, Y^n, Z^n,W^n)$ are assumed to be i.i.d.; i.e. $p(x^n, y^n, z^n,w^n) = \prod_{i=1}^n p(x_i, y_i, z_i,w_i)$. We now give definitions for the case when the switch is opened; i.e. only $Y^n$ is available at the decoder. 
\subsection{Uncoded S.I. case (Figure~\ref{fig1})} \label{sect:2A}
An $(n, 2^{nR})$ code for this setup consists of
\begin{itemize}
\item A \textit{stochastic} encoder $F_e$ that takes $(X^n, Y^n)$ as input and generates $M \in [1:2^{nR}]$ according to a conditional pmf $p(m|x^n, y^n)$; and
\item A decoder $f_D: M\times \Yc^n \to \hat{\Xc}^n$.
\end{itemize}
The \emph{ expected distortion} incurred by the code is given by $\E d(X^n, \Xh^n): = \sum_{i=1}^n E d(X_i, \Xh_i)/n$, where $d: \Xc \times \mathcal{\Xh} \to [0, \infty)$ is the per symbol distortion measure. The \emph{information leakage rate} at the eavesdropper is given by $I(X^n; M, Z^n)/n$. A $(R, D, \Delta)$ R.D.I. tuple is said to be achievable if there exists a sequence of $(n, 2^{nR})$ codes such that
\begin{align}
\limsup_{n \to \infty}\E d(X^n, \Xh^n) \le D, \label{eqn:exp_dist}\\
\limsup_{n \to \infty} \frac{I(X^n; Z^n, M)}{n} \le \Delta. \label{eqn:equivocate}
\end{align}
The \emph{rate-distortion-information leakage rate} (R.D.I.) region is then defined as the closure of all achievable $(R, D, \Delta)$ tuples.

\begin{remark} Another definition of uncertainty used by some authors in the case of discrete memoryless sources is the equivocation rate, defined by $H(X^n|M, Z^n)/n$. Our information leakage rate definition is equivalent to the equivocation rate, since $H(X^n)/n = H(X)$ is fixed.
\end{remark}

\subsection{When $Z^n$ is also available at the decoder} \label{sect:2B}
This setting refers to Figure~\ref{fig1} with the switch closed. When $Z^n$ is also available at the encoder, all of the definitions remain the same with the exception of the decoding function, which is now changed to $f_D: M\times \Yc^n\times \Zc^n \to \hat{\Xc}^n$, since the S.I. $Z^n$ is also available at the decoder.

\subsection{Rate limited helper case (Figure~\ref{fig2})}
The rate limited helper setting is shown in Figure~\ref{fig2}. An $(n, 2^{nR}, 2^{nR_h})$ code for this setup consists of 
\begin{itemize}
\item A stochastic \emph{helper} encoder $F_h$ that takes $Y^n$ as input and outputs $M_h \in [1:2^{nR_h}]$ according to the conditional pmf $p(m_h|y^n)$;
\item A stochastic encoder $F_e$ that takes $(X^n, M_h)$ as input and generates $M \in [1:2^{nR}]$ according to the conditional pmf $p(m|x^n, m_h)$;
\item A decoder $f_D: M\times M_h \times \Zc^n\to \hat{\Xc}^n$.
\end{itemize} 
The definitions of expected distortion incurred by the decoder and information leakage rate at the eavesdropper are the same as previous setting, with $Z^n$ replaced by $W^n$ for the information leakage rate. A $(R, R_h, D, \Delta)$ tuple is said to be achievable if there exists a sequence of $(n, 2^{nR}, 2^{nR_h})$ codes such that~\eqref{eqn:exp_dist} and~\eqref{eqn:equivocate} are satisfied. The R.D.I. region is then defined as the closure of all achievable $(R, R_h, D, \Delta)$ tuples.
\begin{remark}
It should be noted that the rate limited helper setting does not include the previous setting as a special case. The helper encoder is a stochastic encoder. Hence, it can choose to send independent randomness instead of transmitting the $Y^n$ sequence to the encoder and the decoder. 
\end{remark}
\subsection{Side information and rate distortion region}
Let $\Yt^n$ be another i.i.d. random variable such that $(X^n, \Yt^n)\sim \prod_{i=1}^n p(x_i, \yt_i)$. Let $R_{\rm WZ}(D)$ be the rate-distortion function for the Wyner-Ziv setting (see~\cite[Chapter 11]{El-Gamal--Kim2010}) where S.I. $\Yt^n$ is available at the decoder only. Let $R_{\rm SI-Enc}(D)$ be the rate-distortion function when $\Yt^n$ is also available at the encoder. We say that \emph{S.I. at the encoder does not improve the rate-distortion region for side information $\Yt^n$} if $R_{\rm WZ}(D) = R_{\rm SI-Enc}(D)$ for all $D \ge D_{\rm min}$, where $D_{\rm min}$ is the minimum achievable distortion. We denote this condition by $\Rc_{\rm WZ}(\Yt)=\Rc_{\rm SI-Enc}(\Yt)$, where $\Rc_{\rm WZ}(\Yt)$ is the rate distortion region when $\Yt^n$ is available at the decoder only, and $\Rc_{\rm SI-Enc}(\Yt)$ is the rate distortion region when $\Yt^n$ is available at both the encoder and the decoder. Equivalently, we have $\Rc_{\rm WZ}(\Yt) \subseteq \Rc_{\rm SI-Enc}(\Yt)$ and $\Rc_{\rm SI-Enc}(\Yt) \subseteq \Rc_{\rm WZ}(\Yt)$. 

The following information-theoretic characterization of $\Rc_{\rm WZ}(\Yt)=\Rc_{\rm SI-Enc}(\Yt)$ will be useful in the sequel. $\Rc_{\rm WZ}(\Yt)=\Rc_{\rm SI-Enc}(\Yt)$ if for all $D \ge 0$, there exists an auxiliary random variable $V$ and reconstruction function $\xh(V, \Yt)$ such that
\begin{align*}
I(X;V|\Yt) &= R_{\rm SI-Enc}(D) \\
& = \min_{p(\xh|x,\yt): \E d(X, \Xh) \le D} I(X;\Xh|\Yt),
\end{align*}
with $V-X-\Yt$ and $\E d(X, \xh(V,\Yt)) \le D$. 
\section{Uncoded S.I. at encoder and decoder with switch opened} \label{sect:3}
In this section, we present results for the setting in Figure~\ref{fig1} with the switch opened.

\subsection{General inner and outer bounds}
\begin{proposition} \label{prop1}
An outer bound to the R.D.I. region for the setting in Figure~\ref{fig1} with the switch opened is given by
\begin{align*}
R &\ge I(X;U,V|Y), \\
\Delta &\ge \max\left\{\begin{array}{l}I(X;Z), \\ 
I(X;Z,V,U) + I(V;Z|U) \\
- I(V;Y|U) - H(Y|U,V,X,Z)\end{array}\right\},
\end{align*}
for some $p(x,y,z)p(u,v|x,y)$ and reconstruction function $\xh(Y, U,V)$ satisfying $\E d(X, \xh(Y, U,V)) \le D$. The cardinalities of $U$ and $V$ may be upper bounded by $|\Uc| \le |\Xc||\Yc|+2$ and $|\Vc| \le|\Xc||\Yc|+2$.
\end{proposition}
Proof of this Proposition is given in Appendix~\ref{appen:prop1}.

We now present an inner bound (achievability scheme) for this setting. 
\begin{proposition} \label{prop2}
An inner bound to the R.D.I. region for the setting in Figure~\ref{fig1} with the switch opened is given by{\allowdisplaybreaks
\begin{align*}
R &> I(X;U,V|Y), \\
\Delta &> I(X;Z,U) + I(V;X|U,Y) -R_K,
\end{align*}}
where $R_K = \min\{I(V;X|U,Y), H(Y|U,V,X,Z)\}$ for $p(u,v,x,y,z) = p(x,y)p(u,v|x,y)p(z|x,y)$ and reconstruction function $\xh(Y, U,V)$ satisfying $\E d(X, \xh(Y, U,V)) \le D$. 
\end{proposition}

Proof of Proposition~\ref{prop2} is given in Appendix~\ref{appen:prop2}. Here, we give some intuition behind the general achievability scheme. The encoder sends two layers of descriptions $U^n$ and $V^n$ to the decoder, which decodes by successive decoding. This results in rates of $I(X;U|Y)$ for the first $U^n$ layer and $I(V;X|U,Y)$ for the second layer. We assume that the eavesdropper is able to decode the $U^n$ codeword, resulting in side information $(Z^n, U^n)$ at the eavesdropper. S.I. $Y^n$ is binned to $2^{nR_K}$ bins to generated a secret key. This key can be kept secret from the eavesdropper if $R_K \le H(Y|U,V,X,Z)$, and it is then used to scramble the message sent to the decoder about the $V^n$ layer of codewords. This operation increases the uncertainty that the eavesdropper has about the $V^n$ codewords. The information leakage rate is then upper bounded by $I(X;Z,U)$ plus $I(V;X|U,Y) -R_K$. $I(V;X|U,Y)$ is an upper bound on the leakage rate due to the $V^n$ codeword if no scrambling was done, while $-R_K$ represents the reduction in the leakage rate due to the secret key scrambling operation. 

\begin{remark}
The reader may ask why we did not scramble the first layer of codewords. A straightforward way of scrambling the first layer of codewords as well as the second layer is to define in the inner bound $U = \emptyset$ and $V' = (V,U)$. Such a scheme leads to the following R.D.I. trade-off.{\allowdisplaybreaks
\begin{align*}
R &> I(V';X|Y) \\
& = I(V,U; X|Y), \\
\Delta &> I(X;Z) + I(V';X|Y) - R_K \\
& = I(X;Z) + I(V, U;X|Y) - R_K,
\end{align*} }
where $R_K = \min\{I(V;X|U,Y), H(Y|U,V,X,Z)\}$. 
\end{remark}
\begin{remark}
As a sanity check, it is easy to see that if we set $Y = \emptyset$, Propositions~\ref{prop1} and~\ref{prop2} allow us to recover a special case of the result in~\cite{Villard2010}, where S.I. is not available at the encoder. The R.D.I. region in this case is given as
\begin{align*}
R &\ge I(X;V), \\
\Delta &\ge I(X;Z,V) + I(V;Z)
\end{align*}
for some $p(x,y,z)p(v|x,y)$ and reconstruction function $\xh(Y, V)$ satisfying $\E d(X, \xh(Y, V)) \le D$. The cardinality of $V$ may be upper bounded by $|\Vc| \le|\Xc||\Yc|+2$.
\end{remark}

\subsection{R.D.I. regions}
\begin{proposition} \label{prop3}
For the setting in Figure~\ref{fig1} with the switch opened, if $X-Y-Z$ and $\Rc_{\rm SI-Enc}(Y)= \Rc_{\rm WZ}(Y)$, the R.D.I. region is given by
\begin{align*}
R &\ge R_{\rm SI-Enc}(D), \\
\Delta & \ge \max\{I(X;Z), I(X;Z) + R_{\rm SI-Enc}(D) - H(Y|X,Z)\}.
\end{align*}
Here, $R_{\rm SI-Enc}(D) = \min_{p(\xh|x,y): \E d(X, \Xh) \le D}I(X;\Xh|Y)$.
\end{proposition}
Proof of this Proposition follows from tightening the outer bound in Proposition~\ref{prop1} using the two conditions and showing achievability using Proposition~\ref{prop2}.
\begin{IEEEproof}
From Proposition~\ref{prop1}, we have{\allowdisplaybreaks
\begin{align*}
\Delta &\ge I(X;Z,V,U) + I(V;Z|U) - I(V;Y|U) \\
& \qquad - H(Y|U,V,X,Z) \\
& \stackrel{(a)}{\ge} I(X;Z,V,U) + I(V,U;Z) - I(V,U;Y) \\
& \qquad- H(Y|U,V,X,Z) \\
& = I(X;Z) + I(X;V,U|Z)+ I(V,U;Z) - I(V,U;Y) \\
& \qquad - H(Y|U,V,X,Z) \\
& = I(X;Z) + I(X;V,U) - I(V,U;Y) \\
& \qquad - H(Y|U,V,X,Z) + I(Z;V,U|X)\\
& = I(X;Z) + I(X,Y;V,U) - I(Y;V,U|X) - I(V,U;Y) \\
& \qquad - H(Y|X,Z)+ I(Y;V,U|X,Z)+ I(Z;V,U|X)\\
& = I(X;Z) + I(X;V,U|Y) - H(Y|X,Z)\\
& \qquad- I(Y;V,U|X) + I(Z,Y;V,U|X)\\
& = I(X;Z) + I(X;V,U|Y) - H(Y|X,Z) \\
& \stackrel{(b)}{\ge} I(X;Z) + I(X;\Xh|Y) - H(Y|X,Z) \\
& \ge I(X;Z) + R_{\rm SI-Enc}(D) - H(Y|X,Z).
\end{align*}}
$(a)$ follows from the Markov Chain assumption; $(b)$ follows from $\Xh$ being a function of $(V,U,Y)$; the final step follows from the fact that $R_{\rm SI-Enc}(D) = \min_{p(\xh|x), \E d(\Xh,X)}I(X;\Xh|Y)$. Similarly, from Proposition~\ref{prop1}, a lower bound on $R$ is given by
\begin{align*}
R &\ge I(X; V,U|Y) \\
& \ge R_{\rm SI-Enc}(D).
\end{align*}
This completes the proof of converse.

Achievability follows from Proposition~\ref{prop2} and the assumption that $\Rc_{\rm SI-Enc}(Y) = \Rc_{\rm WZ}(Y)$. Since $\Rc_{\rm SI-Enc}(Y) = \Rc_{\rm WZ}(Y)$, there exists a $V^*$ and reconstruction function $x^*(V^*,Y)$ such that $V^* - X-(Y,Z)$, $I(X;V^*|Y) = R_{\rm WZ}(D) = R_{\rm SI-Enc}(D)$ and $\E d(X, \xh^*(V,Y,Z)) \le D$ for all $D \ge D_{\rm min}$. It is now straightforward to verify that the R.D.I. region stated in the Proposition can be achieved by setting $U = \emptyset$, $V = V^*$ and using the Markov relation $V^*-X-(Y,Z)$. 
\end{IEEEproof}
\begin{remark}
The S.I. at the encoder has, in general, dual uses. One use is to allow the encoder to reduce the rate needed to achieve a level of distortion at the decoder, and the other use here is to generate a secret key. There is, in general, a tension between these two uses of the S.I.. The assumption of $\Rc_{\rm SI-Enc}(Y) = \Rc_{\rm WZ}(Y)$ removes some of this tension, allowing us to characterize the R.D.I. region under certain conditions. This is a recurring theme in this paper. 
\end{remark}
\subsection{Examples}
We now provide two examples involving canonical sources and distortion measures in information theory that satisfy the two assumptions stated in the previous subsection.
\begin{corollary} \label{coro1}
Let $X-Y-Z$ and $Y$ be an erased version of $X$. That is $Y = X$ with probability $1-p_e$, and $e$ with probability $p_e$. Let $|\hat{\mathcal{X}}| = |\Xc|$ and the distortion measure be the Hamming distance:
\begin{align*}
d(X, \Xh) =  \left\{\begin{array}{ll} 0 & \mbox{if } \Xh = X \\ 1 & \mbox{if } \Xh \neq X \end{array}\right. .
\end{align*} 
Then, the R.D.I. region is given by
\begin{align*}
R & \ge p_e I(X; \Xh), \\
\Delta & \ge \max\{I(X;Z), I(X;Z) + p_e I(X;\Xh) - H(Y|X,Z)\}
\end{align*}
for $0 \le D \le p_e$, $p(\xh|x)$ such that $\E d(X, \Xh) \le D/p_e$. 
\end{corollary} 
\begin{IEEEproof}
The proof follows from an application of Proposition~\ref{prop3} and a result in~\cite[Theorem 6]{perron}. Since $X-Y-Z$ by assumption, it remains to check that $\Rc_{\rm SI-Enc}(Y) = \Rc_{\rm WZ}(Y)$, which follows from~\cite[Theorem 6]{perron}. Further,~\cite[Theorem 6]{perron} states that $R_{\rm SI-Enc}(D) = p_e\min_{p(\xh|x): \E d(X, \Xh) \le D/p_e}I(X;\Xh)$. \\
\end{IEEEproof}

\begin{corollary} \label{coro2}
Let $X-Y-Z$ and let the distortion measure be given by the log-loss distortion~\cite{Courtade2011}. That is, the reconstruction alphabet is a vector representing the set of probability distributions of the source $X$. Thus, $\xh(x)$, $1 \le x \le |\Xc|$, represents the $x$ component of the vector $\xh$ that gives the estimated probability of $X = x$. Then, the log-loss measure is defined by
\begin{align*}
d(x, \xh) = \log\frac{1}{\xh(x)}.
\end{align*} 
With this distortion measure, the R.D.I. region is given by
\begin{align*}
R & \ge [H(X|Y) - D]^+, \\
\Delta & \ge \max\{I(X;Z), I(X;Z) + H(X|Y) - D - H(Y|X,Z)\}.
\end{align*}
\end{corollary}
\begin{IEEEproof}
This result follows again from a straightforward application of Proposition~\ref{prop3}. The fact that $\Rc_{\rm SI-Enc}(Y) = \Rc_{\rm WZ}(Y)$ for arbitrary discrete memoryless $X,Y$ under logarithmic loss follows from results in~\cite{Courtade2011}. Further,~\cite{Courtade2011} showed that $R_{\rm SI-Enc}(D) = [H(X|Y) - D]^+$. 
\end{IEEEproof}
\begin{remark}
Technically, our proof of achievability in Proposition~\ref{prop2} holds only for bounded distortion measures, and log-loss is not a bounded distortion measure. The proof of achievability can be readily extended to log-loss by perturbing the reconstruction probability distribution, as was done in an earlier version of~\cite{Courtade2012b}. Fix a desired $p(u,v|x,y)$ in Proposition~\ref{prop2}. For every $u\in \Uc$, $v \in \Vc$ and $y \in \Yc$, define $\Xc_1(u,v,y):=\{x: p(x|u,v,y)>0\}$ and $\Xc_0(u,v,y):=\{x: p(x|u,v,y)=0\}$. Further, let $ \e>0$ be a number such that $\e < (1-\e)\min_{u, v,y, x \in \Xc_1(u,v,y)}p(x|u,v,y)$. Then, we define 
\begin{align*}
\xh(x):=\left\{ \begin{array}{ccc}(1-\frac{|\Xc_0|(u,v,y)}{|\Xc|}\e)p(x|u,v,y) & \mbox{for} & x \in \Xc_1 \\ \frac{\e}{|\Xc|} & \mbox{for} & x\in \Xc_0 \end{array}\right..
\end{align*}
It is then easy to see that the maximum distortion we incur is upper bounded by $\log (|\Xc|/\e)$. The proof in Proposition~\ref{prop2} can then be applied with this reconstruction function. Following the proof in Proposition~\ref{prop2}, let $p_e^{(n)}$ be the probability of ``error''; that is, the probability that the chosen codewords are not jointly typical with $(X^n, Y^n)$ or that the decoder makes an error. Then, for $n$ sufficiently large, the expected distortion under log-loss with the chosen reconstruction function is upper bounded by
\begin{align*}
\E d(X^n, \xh^n(U^n, V^n, Y^n)) &\le D + \d(\e) + p_e^{(n)} \log \left(\frac{|\Xc|}{\e}\right).
\end{align*}
Since $p_e^{(n)} \to 0$ as $n \to \infty$, this completes the proof for the case of log-loss.
\end{remark}
\begin{remark} \label{rm1}
For the case of log-loss, by letting $D \to 0$, we can also recover the lossless source coding case. That is, when the criteria at the decoder is the block error probability $\P(\Xh^n \neq X^n) \to 0$ as $n \to \infty$. Proof of this claim follows from Proposition~\ref{prop5} in Section~\ref{sect:5} relating log-loss in this setting to list decoding. 
\end{remark}

\subsection*{Numerical examples for Corollaries~\ref{coro1} and~\ref{coro2}}

\begin{figure}
\psfrag{Erased S.I.}[l]{Corollary~\ref{coro1}}
\psfrag{Log-Loss}[l]{Corollary~\ref{coro2}}
\begin{center}
\scalebox{0.75}{\includegraphics{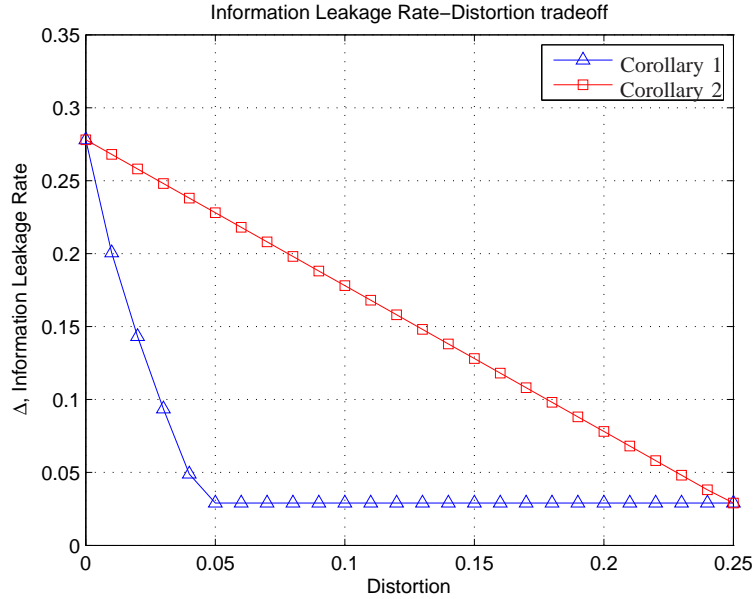}}
\caption{Optimal $\Delta$ and $D$ tradeoff for the numerical examples given for Corollaries~\ref{coro1} and~\ref{coro2}. The blue line with triangles corresponds to the numerical example for Corollary~\ref{coro1}, while the red line with squares corresponds to the numerical example for Corollary~\ref{coro2}.} \label{fig3}
\end{center}
\vspace{-15pt}
\end{figure}

As concrete numerical examples, we consider $X \in \Bern(1/2)$, $p_e = 0.8$ and $Z \in \{0,1\}$ with $\P(Z = 0|Y = 0) = 1$, $\P(Z = 1|Y = 1) = 1$ and $\P(Z= 0|Y = e) = 0.5$. We then have the following R.D.I. regions for the two corollaries.
\begin{enumerate}
\item Numerical example for Corollary~\ref{coro1}: The R.D.I. region is given by
\begin{align*}
R & \ge p_e \left(1 - H_2\left(\frac{D}{p_e}\right)\right), \\
\Delta & \ge \max\left\{1 - H_2(p_e/2), 1 - H_2(p_e/2) + p_e \left(1 - H_2\left(\frac{D}{p_e}\right)\right) - (1-\frac{p_e}{2})H_2\left(\frac{0.5p_e}{1-\frac{p_e}{2}}\right) \right\}
\end{align*}
for $D \le p_e/2$. For $D > p_e/2$, $R = 0$ and $\Delta = 1 - H_2(p_e/2)$. Here, $H_2(.)$ represents the binary entropy function. 

\item Numerical example for Corollary~\ref{coro2}: The R.D.I. region is given by
\begin{align*}
R & \ge [p_e - D]^+, \\
\Delta & \ge \max\left\{1 - H_2(p_e/2), 1 - H_2(p_e/2) + p_e-D - (1-\frac{p_e}{2})H_2\left(\frac{0.5p_e}{1-\frac{p_e}{2}}\right)\right\}
\end{align*}
for $D \ge 0$.
\end{enumerate}
The optimal information leakage rate-distortion tradeoffs for both examples are plotted in Fig.~\ref{fig3}.

\section{Uncoded S.I. at encoder and decoder with switch closed} \label{sect:4}
We now turn our attention to the case where the eavesdropper's side information is also available at the decoder. We note here that this setting is closely related to the setting considered in the previous section. However, this setting cannot be recovered as a special case of the setting in the previous section. One cannot, for example, define $\Yt = (Y,Z)$ as a super-source since that would mean that the eavesdropper's side information would also be available at the encoder. Using the results of this section and the previous section, we show that when $\Rc_{\rm WZ}(Y,Z)=\Rc_{\rm SI-Enc}(Y,Z)$, knowledge of the eavesdropper's side information at the encoder does not change the R.D.I. region (for the setting in Figure~\ref{fig1} with the switch closed). 

\subsection{Inner and outer bounds} \label{sect:4a}

We first start with an inner bound.
\begin{proposition} \label{propyz_in}
An inner bound to the R.D.I. region for the setting in Figure~\ref{fig1} with the switch closed is given by
\begin{align*}
R &> I(X;U,V|Y,Z), \\
\Delta &> I(X;Z,U) + I(V;X|U,Y, Z) -R_K,
\end{align*}
where $R_K = \min\{I(V;X|U,Y, Z), H(Y|U,V,X,Z)\}$ for $p(u,v)p(u,v|x,y)$ and reconstruction function $\xh(Y, Z, U,V)$ satisfying $\E d(X, \xh(Y,Z, U,V)) \le D$. 
\end{proposition}
We omit the proof of this proposition here, as the achievability scheme is largely similar to the achievability scheme of Proposition~\ref{prop2}, with the difference being that the decoder has access to side informations $(Y^n, Z^n)$. Hence, the decoder uses the side informations $Y^n$ and $Z^n$ in decoding the codeword from the encoder, as opposed to just using the side information $Y^n$. Similarly, $Z^n$ is also used in the reconstruction. The rest of the achievability scheme follows the same steps as that in Proposition~\ref{prop2}.

Next, we turn to an outer bound for this setting. 
\begin{proposition} \label{propyz_out}
An outer bound to the R.D.I. region for the setting in Figure~\ref{fig1} with the switch closed is given by
\begin{align*}
R &\ge I(X;V|Y,Z), \\
\Delta &\ge \max\left\{I(X;Z),I(X;Z) + I(X;V|Y,Z) - H(Y|X,Z)\right\},
\end{align*}
for some $p(x,y,z)p(v|x,y)$ and reconstruction function $\xh(Y, Z, V)$ satisfying $\E d(X, \xh(Y, Z, V)) \le D$. The cardinality of $V$ may be upper bounded by $|\Vc| \le|\Xc||\Yc|+2$.
\end{proposition}
Proof of this Proposition is given in Appendix~\ref{appen:propyz_out}.
\subsection{R.D.I. regions}
Using Propositions~\ref{propyz_in} and~\ref{propyz_out}, we characterize the R.D.I. regions for sources and distortion measures satisfying $\Rc_{\rm WZ}(Y,Z) = \Rc_{\rm SI-Enc}(Y,Z)$.
\begin{proposition} \label{propyz_rdi}
For the setting in  Figure~\ref{fig1} with the switch closed, when $\Rc_{\rm WZ}(Y,Z) = \Rc_{\rm SI-Enc}(Y,Z)$, the R.D.I. region is given by{\allowdisplaybreaks
\begin{align*}
R &\ge R_{\rm SI-Enc}(D), \\
\Delta &\ge \max\left\{I(X;Z),I(X;Z) + R_{\rm SI-Enc}(D) - H(Y|X,Z)\right\}.
\end{align*}}
Here, $R_{\rm SI-Enc}(D) = \min_{p(\xh|x,y,z): \E d(X, \Xh) \le D}I(X;\Xh|Y,Z)$.\\
\end{proposition}
\begin{IEEEproof}
From the outer bound in Proposition~\ref{propyz_out}, we have
\begin{align*}
R &\ge I(X;V|Y,Z) \\
& \ge I(X;\Xh|Y,Z) \\
& \ge R_{\rm SI-Enc}(D).
\end{align*}
Similarly, we have
\begin{align*}
\Delta &\ge \max\left\{I(X;Z),I(X;Z) + R_{\rm WZ}(D) - H(Y|X,Z)\right\}.
\end{align*}
Achievability of this outer bound then follows from Proposition~\ref{propyz_in} and the assumption that $\Rc_{\rm SI-Enc}(Y,Z) = \Rc_{\rm WZ}(Y,Z)$. Since $\Rc_{\rm SI-Enc}(Y,Z) = \Rc_{\rm WZ}(Y,Z)$, there exists a $V^*$ and reconstruction function $x^*(V^*,Y,Z)$ such that $V^* - X-(Y,Z)$, $I(X;V^*|Y,Z) = R_{\rm WZ}(D) = R_{\rm SI-Enc}(D)$ and $\E d(X, \xh^*(V,Y,Z)) \le D$ for all $D \ge D_{\rm min}$. We then set $U = \emptyset$ and $V = V^*$ in the inner bound in Proposition~\ref{propyz_in} to show the achievability of the outer bound.
\end{IEEEproof}

Under the condition that $\Rc_{\rm WZ}(Y,Z) = \Rc_{\rm SI-Enc}(Y,Z)$, Proposition~\ref{propyz_rdi} and Proposition~\ref{prop3} allow us to show that the R.D.I. region of the setting in Figure~\ref{fig1} does not change even if the eavesdropper's S.I. is available to both the encoder and the decoder. This is stated in the next proposition.
\begin{proposition} \label{prop_rdi_all}
For the setting in Figure~\ref{fig1} with the switch closed, if $\Rc_{\rm WZ}(Y,Z) = \Rc_{\rm SI-Enc}(Y,Z)$, the R.D.I. region remains unchanged even if $Z^n$ is available at the encoder.
\end{proposition}
\begin{IEEEproof}
Proof of this Proposition follows quite straightforwardly from Proposition~\ref{prop3}. We let the side information observed by the decoder be the super source $\Yt^n = (Y^n,Z^n)$. Observe that since $X-\Yt - Z$ and $\Rc_{\rm WZ}(Y,Z) = \Rc_{\rm SI-Enc}(Y,Z)$ implies that $\Rc_{\rm WZ}(\Yt) = \Rc_{\rm SI-Enc}(\Yt)$, the results of Proposition~\ref{prop3} holds and the eavesdropper's S.I. $Z^n$ now becomes available to both the encoder and the decoder. It is now straightforward to see from Proposition~\ref{prop3} that the R.D.I. region is the same as that given in Proposition~\ref{propyz_rdi}. 
\end{IEEEproof}
\subsection{Examples}
We now give examples of sources and distortion measures satisfying the condition $\Rc_{\rm WZ}(Y,Z) = \Rc_{\rm SI-Enc}(Y,Z)$.

\begin{corollary} \label{coro3}
Let $X-Z-Y$ and $Z$ be an erased version of $X$. That is $Z = X$ with probability $1-p_e$, and $e$ with probability $p_e$. Let $|\hat{\mathcal{X}}| = |\Xc|$ and the distortion measure be the Hamming distance, as defined in Corollary~\ref{coro1}.
Then, the R.D.I. region is given by
\begin{align*}
R & \ge p_e I(X; \Xh), \\
\Delta & \ge \max\{I(X;Z), I(X;Z) + p_e I(X;\Xh) - H(Y|Z)\}
\end{align*}
\end{corollary} 
for $0 \le D \le p_e$, $p(\xh|x)$ such that $\E d(X, \Xh) \le D/p_e$. 
\begin{IEEEproof}
Proof of this Corollary follows similar lines to that of Corollary~\ref{coro1}. However, we first show that knowledge of S.I. $Y^n$ at both the encoder and the decoder does not improve the rate-distortion region, when S.I. $Z^n$ is also known at the encoder and decoder and $X-Z-Y$ form a Markov Chain. When S.I.s $Z^n$ and $Y^n$ are known at both the encoder and the decoder, the rate distortion function, $R_{\rm SI-Enc}(D)$, is given as
\begin{align*}
R_{\rm SI-Enc}(D) &= \min I(X;\Xh|Y,Z),
\end{align*}
where the minimization is over $p(\xh|x,y,z)$ satisfying $\E d(X, \Xh) \le D$. Note now that using the Markov Chain $X-Z-Y$, we have that $I(X;\Xh|Y,Z) \ge I(X; \Xh|Z)$. Since $I(X;\Xh|Z)$ and $\E d(X, \Xh)$ depend on only the marginal p.m.f. $p(x,z,\xh)$, the rate distortion function can be equivalently written as
\begin{align*}
R_{\rm SI-Enc}(D) &= \min_{p(\xh|x,z): \E d(X, \Xh) \le D} I(X;\Xh|Z).
\end{align*}
Hence, S.I. $Y^n$ does not improve the rate-distortion region when the Markov Chain $X-Z-Y$ holds, and we have $\Rc_{\rm SI-Enc}(Y,Z) = \Rc_{\rm SI-Enc}(Z)$. 

Using the result in~\cite[Theorem 6]{perron}, we have $\Rc_{\rm SI-Enc}(Z) = \Rc_{\rm WZ}(Z)$. Next, noting that $\Rc_{\rm WZ}(Z) \subseteq \Rc_{\rm WZ}(Y,Z) \subseteq \Rc_{\rm SI-Enc}(Y,Z)$ then give us the required condition $\Rc_{\rm SI-Enc}(Y, Z) = \Rc_{\rm WZ}(Y, Z)$.

Finally, we apply Proposition~\ref{propyz_rdi} and~\cite[Theorem 6]{perron} to obtain the R.D.I. region in Corollary~\ref{coro3}.   
\end{IEEEproof}

\begin{remark}
In this example, the eavesdropper's S.I., $Z^n$, is of higher quality than the S.I. observed by the encoder and decoder, $Y^n$. $Y^n$ therefore plays no role in reducing the achievable rate for a given distortion. However, because $Y^n$ is observed at both the encoder and decoder, it can still help to reduce the information leakage rate, despite it being a degraded version of $Z^n$.
\end{remark}
Our next example deals with the case where both $Y$ and $Z$ are erased versions of $X$.
\begin{corollary} \label{coro4}
Let $Y$ be an erased version of $X$. That is $Y = X$ with probability $1-p_{e,y}$, and $e$ with probability $p_{e,y}$. Similarly, let $Z$ be an erased version of $X$, independent of $Y$ conditioned on $X$. That is $Z = X$ with probability $1-p_{e,z}$, and $e$ with probability $p_{e,z}$. Let $|\hat{\mathcal{X}}| = |\Xc|$ and the distortion measure be the Hamming distance as defined in Corollary~\ref{coro1}.
Then, the R.D.I. region is given by
\begin{align*}
R & \ge p_{e,y}p_{e,z} I(X; \Xh), \\
\Delta & \ge \max\{I(X;Z), I(X;Z) + p_{e,y}p_{e,z} I(X;\Xh) - H(Y|X)\}
\end{align*}
for $0 \le D \le p_{e,y}p_{e,z}$, $p(\xh|x)$ such that $\E d(X, \Xh) \le D/(p_{e,y}p_{e,z})$. 
\end{corollary} 
\begin{IEEEproof}
Similar to Corollary~\ref{coro3}, we use Proposition~\ref{propyz_rdi} to prove this result. It remains to check that $\Rc_{\rm SI-Enc}(Y,Z) = \Rc_{\rm WZ}(Y,Z)$ when $Y$ and $Z$ are both erased versions of $X$. This fact is a straightforward extension of the arguments in~\cite[Theorem 6]{perron}. We therefore omit it here. 
\end{IEEEproof}
\begin{remark}
It may be of interest to compare Corollary~\ref{coro4} to the setting in Figure~\ref{fig1} when the switch is opened, with the side information at the decoder being replaced by the following erased side information: $\Yt = X$ with probability $1 - p_{e,y}p_{e,z}$ and $e$ with probability $p_{e,y}p_{e,z}$, and $X-\Yt-Z$. In this case, from Corollary~\ref{coro1}, the R.D.I. region is given by 
\begin{align*}
R_{\rm open} & \ge p_{e,y}p_{e,z} I(X; \Xh), \\
\Delta_{\rm open} & \ge \max\{I(X;Z), I(X;Z) + p_{e,y}p_{e,z} I(X;\Xh) - H(\Yt|X, Z)\}
\end{align*}
for $0 \le D \le p_{e,y}p_{e,z}$, $p(\xh|x)$ such that $\E d(X, \Xh) \le D/(p_{e,y}p_{e,z})$. In this case, the expression for $R_{\rm open}$ is the same as that for $R$ in Corollary~\ref{coro4}. This is to be expected since, for rate distortion, observing two erased side informations $Y$ and $Z$ is equivalent to observing a higher quality erased side information $\Yt$. However, the information leakage rate expressions are different, since $H(\Yt|X,Z)$ is in general not equal to $H(Y|Z)$. Hence, due to the required Markov Chain assumption ($X-\Yt-Z$) in Corollary~\ref{coro1}, the result in Corollary~\ref{coro4} cannot be recovered from Corollary~\ref{coro1} by simply assuming a higher quality erased side information at the decoder. 
\end{remark}

Our final example deals with the setting under log-loss.

\begin{corollary} \label{coro5}
For the setting in Figure~\ref{fig1} with the switch closed, let the distortion measure be given by the log-loss distortion as defined in Corollary~\ref{coro2}. 
The R.D.I. region is given by
\begin{align*}
R & \ge [H(X|Y,Z) - D]^+, \\
\Delta & \ge \max\{I(X;Z), I(X;Z) + H(X|Y,Z) - D - H(Y|X,Z)\},
\end{align*}
where $[x]^+ := \max\{0, x\}$. 
\end{corollary}
\begin{IEEEproof}
The proof follows similar lines to the proof in Corollary~\ref{coro2}, with the role of Proposition~\ref{prop3} being replaced by Proposition~\ref{propyz_rdi}. The fact that $\Rc_{\rm SI-Enc}(Y,Z) = \Rc_{\rm WZ}(Y,Z)$ follows again from results in~\cite{Courtade2011}, by consider $(Y,Z)$ as a super source $\Yt$. Further, using the results in~\cite{Courtade2011}, we have $R_{\rm SI-Enc}(D) = [H(X|Y,Z) - D]^+$.
\end{IEEEproof}
\subsection*{Numerical examples for Corollaries~\ref{coro3},~\ref{coro4} and~\ref{coro5}}
\begin{figure}
\psfrag{Corollary 1}[l]{Corollary~\ref{coro3}}
\psfrag{Corollary 2}[l]{Corollary~\ref{coro4}}
\psfrag{Corollary 3}[l]{Corollary~\ref{coro5}}
\begin{center}
\scalebox{0.75}{\includegraphics{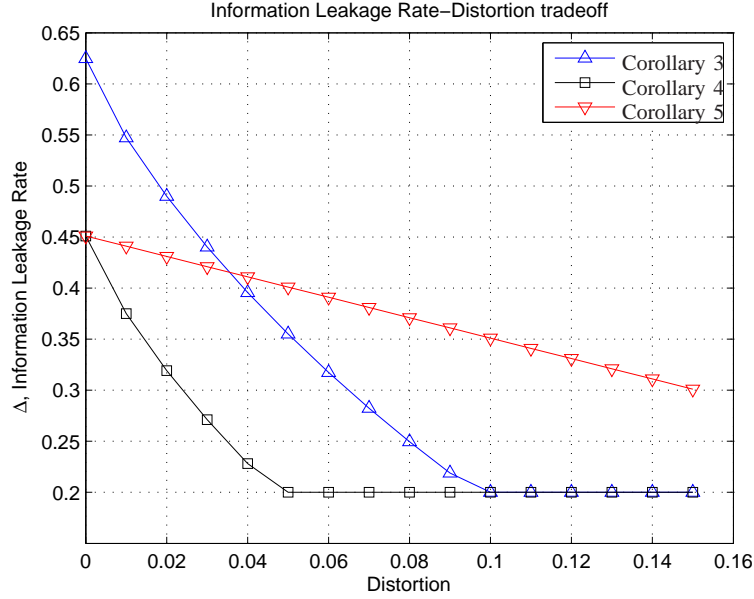}}
\caption{Optimal $\Delta$ and $D$ tradeoff for the numerical examples given for Corollaries~\ref{coro3},~\ref{coro4} and~\ref{coro5}. The blue line with up triangles corresponds to the numerical example for Corollary~\ref{coro3}; the black line with squares corresponds to the numerical example for Corollary~\ref{coro4}, and the red line with down triangles corresponds to the numerical example for Corollary~\ref{coro5}.} \label{fig4}
\end{center}
\vspace{-15pt}
\end{figure}

We now give numerical examples for the three corollaries. For all three examples, we assume that $X\sim \Bern(0.5)$. 
\begin{enumerate}
\item Numerical example for Corollary~\ref{coro3}: We let $Z = X$ with probability $1-p_e$ and $e$ with probability $p_e$, with $p_e = 0.8$. $Y \in \{0,1\}$ with $\P(Y = 0|Z = 0) = 1$, $\P(Y = 1|Z = 1) = 1$ and $\P(Y= 0|Z = e) = 0.9$. The R.D.I. region is given by {\allowdisplaybreaks
\begin{align*}
R & \ge p_e \left(1 - H_2\left(\frac{D}{p_e}\right)\right), \\
\Delta & \ge \max\left\{1-p_e,  1-p_e+ p_e \left(1 - H_2\left(\frac{D}{p_e}\right)\right) - p_eH_2(0.9)\right\}
\end{align*}}
for $D \le p_e/2$. $R = 0$ and $\Delta = 1-p_e$ for $D > p_e/2$.

\item Numerical example for Corollary~\ref{coro4}: We let $Z = X$ with probability $1-p_{e,z}$ and $e$ with probability $p_{e,z}$, with $p_{e,z} = 0.8$. We let $Y = X$ with probability $1-p_{e,y}$ and $e$ with probability $p_{e,y}$, with $p_{e,y} = 0.9$. The R.D.I. region is given by
\begin{align*}
R & \ge p_{e,y}p_{e,z}\left(1 - H_2\left(\frac{D}{p_{e,y}p_{e,z}}\right)\right) , \\
\Delta & \ge \max\left\{1-p_{e,z}, 1-p_{e,z} + p_{e,y}p_{e,z}\left(1 - H_2\left(\frac{D}{p_{e,y}p_{e,z}}\right)\right) - H_2(p_{e,y})\right\}
\end{align*}
for $D \le p_{e,y}p_{e,z}/2$. $R = 0$ and $\Delta = 1-p_{e,z}$ for $D > p_{e,y}p_{e,z}/2$.

\item Numerical example for Corollary~\ref{coro5}: We let $Z = X$ with probability $1-p_{e,z}$ and $e$ with probability $p_{e,z}$, with $p_{e,z} = 0.8$. We let $Y = X$ with probability $1-p_{e,y}$ and $e$ with probability $p_{e,y}$, with $p_{e,y} = 0.9$. The R.D.I. region under log-loss is given by
\begin{align*}
R & \ge [p_{e,z}p_{e,y} - D]^+, \\
\Delta & \ge \max\{1-p_{e,z}, 1-p_{e,z} + p_{e,z}p_{e,y} - D - H_2(p_{e,y})\},
\end{align*}
\end{enumerate}
The optimal information leakage rate-distortion tradeoffs for all three examples are plotted in Fig.~\ref{fig4}.

\section{Rate-limited helper setting} \label{sect:5}
In this section, we consider the rate-limited helper setting in Figure~\ref{fig2}. 
\subsection{General inner bound}
\begin{proposition}  \label{prop4}
An inner bound to the R.D.I. region for the rate limited helper setting in Figure~\ref{fig2} is given by
\begin{align*}
R_h &> \max\{I(U_h;Y|Z), I(U_h;Y|X)\}, \\
R & > I(X;V,U|Z, U_h), \\
\Delta &> I(X;W, U) + I(X;V|Z, U_h, U)\\
& \quad \quad + I(V,U; U_h|X, Y) + I(U; U_h|X,Y)- R_K -R_K'
\end{align*}
for $p(u_h, u,v,x,y,z,w)$ and reconstruction function $\xh(U_h, U,V,Z)$ such that $\E d(X, \xh(U_h, U,V,Z)) \le D$, $R_K \le I(U_h;Y) - I(U_h; X,W,U,V)$, $R_K' \le R_h -\max\{I(U_h;Y|Z), I(U_h;Y|X)\}$, and $R_K +R_K'\le  I(X;V|Z, U_h, U)$. In addition, $p(u_h, u,v,x,y,z,w)$ obey the Markov relations $U_h-Y-(X,Z,W)$, $(V,U)- (X,U_h) - (Y,Z,W)$ and $(V,U,U_h) - (X,Y) - (W,Z)$. That is, $p(u_h, u,v,x,y,z,w) = p(x,y)p(u_h|y)p(v,u|x,u_h)p(w,z|x,y)$.
\end{proposition}

Proof of this proposition is given in Appendix~\ref{rhach}. Here, we give an outline of the proof. The proof follows similar lines to that in Proposition~\ref{prop2}, with the encoder sending two layers of descriptions $U^n$ and $V^n$ to the decoder. The main differences are in the actions of the helper and how the secret key is being generated. To reduce $R$, the helper sends a description $U_h^n$ to both the encoder and the decoder. To ensure that both the encoder and the decoder can decode $U_h^n$, we require $R_h \ge \max\{I(U_h;Y|Z), I(U_h;Y|X)\} $. The secret key is generated in two parts. The first part of the secret key comes from the codeword $U_h^n$. A secret key of rate $R_K$ can be generated by random binning of the $U_h^n$ codewords if $R_K \le I(U_h;Y) - I(U_h; X,W,V,U)$. Next, the helper can also use its own randomness and the remaining rate ($R_K' \le R_h - \max\{I(U_h;Y|Z), I(U_h;Y|X)\}$) to send to the encoder and the decoder a uniform random variable of size up to $2^{n(R_h - \max\{I(U_h;Y|Z), I(U_h;Y|X))\}}$ as a second secret key. Hence, $R_K' \le R_h - \max\{I(U_h;Y|Z), I(U_h;Y|X)\}$. These two keys are then used to scramble the message sent on the rate limited link about the second layer of description $V^n$, which is of rate $I(X;V|U_h, Z,U)$, resulting in the requirement that $R_K +R_K'\le  I(X;V|Z, U_h, U)$. 


In this achievability scheme, there is a tradeoff between the amount of secret key generated and the quality of the description that the helper sends to reduce the rate required by the encoder. The independent randomness sent on the helper link reduces the amount of information leakage through secret key scrambling, but does not help to reduce the distortion at the decoder. While we can generate another secret key using the helper codeword, $U_h^n$, the rate of the key that can be generated is usually not as large as it would be if uniform randomness is used. In some cases such as those in the next subsection, the tradeoff is tight.
\subsection{R.D.I. regions for discrete memoryless source and S.I.s}
We now consider some special cases in which the achievability scheme in Proposition~\ref{prop4} is optimal. 
\begin{proposition} \label{prop6}
For the setting in Figure~\ref{fig2}, if $Y-X-Z-W$ and the distortion measure is log-loss distortion (see definition in Corollary~\ref{coro2}), then the R.D.I. region is given by {\allowdisplaybreaks
\begin{align*}
R_h &\ge I(U_h;Y|Z), \\
R & \ge [H(X|U_h, Z) - D]^+, \\
\Delta &\ge \max\{I(X;W), I(X;W) + H(X|Z)  - D - R_h\}
\end{align*}}
for $p(u_h|y)p(x,z,w|y)$, with $|\Uc_h| \le |\Yc| +2$. 
\end{proposition}
This result generalizes some of the results found in~\cite{Tandon2009}. By setting $W = \emptyset$ and $D = 0$\footnote{See Remark~\ref{rm1} and Proposition~\ref{prop5}}, we recover~\cite[Theorem 4]{Tandon2009} and by setting $Z = \emptyset$ as well, we recover~\cite[Theorem 2]{Tandon2009}.
\begin{IEEEproof}
Achievability of the R.D.I. region in Proposition~\ref{prop6} for $D \le H(X|U_h, Z)$ follows from Proposition~\ref{prop4} by setting $U = \emptyset$, $V$ to be the following random variable
\begin{align*}
V = \left\{\begin{array}{ll}X & \mbox{with probability } 1- \frac{D}{H(X|U_h,Z)} \\ \emptyset & \mbox{otherwise}\end{array} \right. .
\end{align*}
The reconstruction function is given by $\xh(u_h, v, z) :=p(x|u_h,v,z)$ and it can be verified that this reconstruction function achieves $\E d(X, \Xh) = H(X|U_h, V,Z) = D$. 

Next, we note now that the definition of $V$ results in the Markov Chain $V-X-(U_h, Y,Z,W)$. Further, since $Y-X-Z-W$, we have $I(U_h;Y|Z) \ge I(U_h;Y|X)$. The achievable leakage rate is then given by
\begin{align*}
\Delta > I(X;W) + H(X|Z,U_h) -D- R_K -R_K'
\end{align*}
for $R_K \le I(U_h; Y) - I(U_h;X)$, $R_K' \le R_h - I(U_h;Y|Z)$ and $R_K + R_K' \le I(V;X|Z,U_h)$. Hence, the achievable $\Delta$ is either $I(X;W)$, or $ I(X;W) + H(X|Z, U_h)- D- (R_h-I(U_h;X|Z)) = I(X;W) + H(X|Z)  - D - R_h$ if $R_h - I(U_h;X|Z) < H(X|Z,U_h) -D$.

For the proof of the converse, the identification of the auxiliary random variable $U_h$ and lower bounds for the rates $R$ and $R_h$ follow steps similar to those in~\cite{Permuter--Steinberg--Weissman2010}. Further, we will use the following lemma for log-loss found in~\cite{Courtade2012b}. 
\begin{lemma}\label{lemld}
Suppose $\E d(X^n, \Xh^n) \le D$ under log-loss. Then, 
\begin{align*}
H(X^n|Z^n, M, M_h) \le nD.
\end{align*}
\end{lemma}

Given an $(n, 2^{nR}, 2^{nR_h})$ code that achieves $(D+ \e_n, \Delta+\e_n)$, define $U_{h,i}:= (M_h, X^{i-1}, Z^{i-1}, Z_{i+1}^n)$. Note that $U_{h,i}- Y_i-(X_i,Z_i,W_i)$ form a Markov Chain. We have{\allowdisplaybreaks
\begin{align*}
nR_h &\ge H(M_h) \\
& \ge I(Y^n;M_h|Z^n) \\
& = \sum_{i=1}^n I(Y_i; M_h|Z^n, Y^{i-1}) \\
& = \sum_{i=1}^n I(Y_i; M_h|Z^n, Y^{i-1}) \\
& = \sum_{i=1}^n I(Y_i; M_h, Z^{i-1}, Z_{i+1}^n, Y^{i-1}|Z_i) \\
& \stackrel{(a)}{=} \sum_{i=1}^n I(Y_i; M_h, Z^{i-1}, Z_{i+1}^n, X^{i-1}, Y^{i-1}|Z_i) \\
& \ge \sum_{i=1}^n I(Y_i; U_{h,i}|Z_i).
\end{align*} }
$(a)$ follows from the Markov chain $X^{i-1} - (Y^{i-1}, Z^{n}, M_h) - Y_i$, which can be readily shown using techniques in~\cite{Permuter--Steinberg--Weissman2010}.{\allowdisplaybreaks
\begin{align*}
nR &\ge H(M) \\
& \ge I(X^n;M|Z^n, M_h) \\
& = \sum_{i=1}^n I(X_i;M|X^{i-1}, Z^n, M_h) \\
& = \sum_{i=1}^n H(X_i|U_{h,i}, Z_i) - H(X^n|Z^n, M_h, M) \\
& \ge \sum_{i=1}^n H(X_i|U_{h,i}, Z_i) - nD - n\e_n.
\end{align*} }
The last step follows from an application of Lemma~\ref{lemld}. 

For the information leakage term, we have
\begin{align}
 n\Delta + nR_h + \e_n & = I(X^n; M, W^n) + H(M_h) \nonumber\\
& = I(X^n;W^n) + I(X^n;M|W^n)+ H(M_h) \nonumber\\
& \stackrel{(a)}{\ge} I(X^n;W^n) + I(X^n;M|Z^n)+ H(M_h) \nonumber\\
& \ge I(X^n;W^n) + I(X^n;M, M_h|Z^n) - I(X^n; M_h|M,Z^n)+ H(M_h|M, Z^n) \nonumber\\
& \ge I(X^n;W^n) + I(X^n;M, M_h|Z^n). \label{ineq1}
\end{align}
$(a)$ follows from the Markov Chain assumption $Y^n-X^n-Z^n-W^n$; i.e.{\allowdisplaybreaks
\begin{align*}
I(X^n;M|W^n) & = I(Z^n, X^n;M|W^n) - I(Z^n; M|X^n, W^n) \\
& = I(Z^n;M|W^n) + I( X^n;M|Z^n, W^n) \\
& \ge I(X^n; M|Z^n) - I(X^n;W^n|Z^n) \\
& = I(X^n; M|Z^n). 
\end{align*}}
Now, we use Lemma~\ref{lemld} again on the term $H(X^n|M, M_h, Z^n)$ to obtain $H(X^n|M, M_h, Z^n) \le n D -n\e_n$. Hence,{\allowdisplaybreaks
\begin{align*}
n\Delta + nR_h &\ge \sum_{i=1}^n(I(X_i;W_i) + H(X_i|Z_i)) - nD- 2n\e_n. \\
\end{align*}}
The lower bound $n\Delta \ge \sum_{i=1}^nI(X_i;W_i)$ is easy to show.

Now, define $Q\sim \U[1:n]$ independent of all other random variables, and $U_{h} = (Q,U_{h,Q})$, $X_Q = X$, $Y_Q = Y$, $Z_Q = Z$ and $W_Q = W$. It is straightforward to verify that $U_h - Y-(X,Z,W)$ form a Markov Chain. Noting that $\e_n \to 0$ as $n \to \infty$, we arrive at the required bound stated in the Proposition. The cardinality bound on $U_h$ follows from standard techniques~\cite[Appendix C]{El-Gamal--Kim2010}. 
\end{IEEEproof}
The next result presents another case in which Proposition~\ref{prop4} is optimal under a different Markov Chain condition, and for a class of distortion measures that include log-loss. 
\begin{proposition} \label{coro6}
For the setting in Figure~\ref{fig2}, if $Y-W-Z-X$ and $\Rc_{\rm SI-Enc}(Z) = \Rc_{\rm WZ}(Z)$, then the R.D.I. region is given by {\allowdisplaybreaks
\begin{align*}
R_h &\ge 0, \\
R & \ge R_{\rm SI-Enc}(D), \\
\Delta &\ge \max\{I(X;W), I(X;W) + R_{\rm SI-Enc}(D) - R_h\}.
\end{align*}  }
Here, $R_{\rm SI-Enc}(D) = \min_{p(\xh|x,z): \E d(X, \Xh) \le D}I(X;\Xh|Z)$.
\end{proposition}
Proof of this proposition is given in Appendix~\ref{appen:c}. 
In this setting, side information at the decoder is of higher quality than the side information at the encoder. Since we assume that $\Rc_{\rm WZ} = \Rc(\rm SI-Enc)$, any side information sent by the helper does not help to reduce the rate required to achieve a required distortion at the decoder. The helper's only role is to generate a secret key to reduce the information leakage rate. Hence, in this case, there is no tradeoff in the role of the helper between sending a higher quality description versus sending a secret key to reduce the information leakage rate.

\begin{remark}
It may be of interest to note that the achievability scheme in this proposition relies on a helper with enough independent randomness to generate a secret key of size $2^{nR_h}$. The side information $Y^n$ is completely ignored. If, however, the helper is stochastically constrained, in the sense of~\cite{Watanabe2012cf}, then $Y^n$ may be used to generate an additional secret key. A complete characterization of the R.D.I. region for the case of a stochastically constrained helper is, however, an open question to the best of our knowledge.
\end{remark}

Using Proposition~\ref{coro6}, we have the following two examples for erased side information and Hamming distortion, and log-loss distortion.

\begin{corollary}\label{coro7}
For the setting in Figure~\ref{fig2}, if $Y-W-Z-X$, $Z = X$ with probability $1 - p_e$ and $Z = e$ with probability $p_e$ and the distortion measure is Hamming distortion, then the R.D.I. region is given by 
\begin{align*}
R_h &\ge 0, \\
R & \ge p_e I(X;\Xh), \\
\Delta &\ge \max\{I(X;W), I(X;W) + p_e I(X;\Xh) - R_h\},
\end{align*}  
for $p(\xh|x)$ satisfying $\E d(X, \Xh) \le D/p_e$. 
\end{corollary}
\begin{IEEEproof}
The proof follows straightforwardly from Proposition~\ref{coro6}. The fact that $\Rc_{\rm SI-Enc}(Z) = \Rc_{\rm WZ}(Z)$ and $R_{\rm  SI-Enc}(D) =\min_{p(\xh|x): \E d(X, \Xh) \le D/p_e} p_e I(X;\Xh)$ follow from~\cite{perron}.\\
\end{IEEEproof}

\begin{corollary}\label{coro8}
For the setting in Figure~\ref{fig2}, if $Y-W-Z-X$ and the distortion measure is log-loss distortion (see definition in Corollary~\ref{coro2}), then the R.D.I. region is given by{\allowdisplaybreaks 
\begin{align*}
R_h &\ge 0, \\
R & \ge \left[H(X|Z) - D\right]^+, \\
\Delta &\ge \max\{I(X;W), I(X;W) + H(X|Z) - D - R_h\}.
\end{align*}  }
\end{corollary}
\begin{IEEEproof}
The proof again follows straightforwardly from Proposition~\ref{coro6}. The fact that $\Rc_{\rm SI-Enc}(Z) = \Rc_{\rm WZ}(Z)$ and $R_{\rm SI-Enc}(D) =H(X|Z) - D$ follows from~\cite{Courtade2011}.
\end{IEEEproof}

\subsection{Quadratic Gaussian setting}
Following the approach in~\cite{Wyner78} (see also~\cite{Permuter--Steinberg--Weissman2010}), we can extend this setting and analysis to the Quadratic Gaussian case. In this subsection, we consider the sources as zero mean Gaussian sources satisfying the Markov Chain assumption, and the distortion measure is given by the squared distortion measure.

In a close analog to the case of Proposition~\ref{prop6} for log-loss, we have the following result for the Quadratic Gaussian setting.
\begin{proposition} \label{prop8}
For the setting in Figure~\ref{fig2}, let $W\sim N(0, \sigma^2_W)$, $Z =  W + A$, $X = Z+ B$ and $Y = X + C$, where $A \sim N(0, \sigma_A^2)$, $A \sim N(0, \sigma_B^2)$ and $C \sim N(0, \sigma_C^2)$ are mutually independent. To avoid degenerate cases, we assume that $\sigma^2_W, \sigma^2_A, \sigma^2_B, \sigma^2_C > 0$. Let the distortion measure be the squared distortion $d(x, \xh) := (x - \xh)^2$. Then, for fixed $R_h$ and $D$, the R.D.I. region is given by
\begin{align*}
R &\ge \left[\frac{1}{2}\log\left(\frac{\sigma_{B}^2\left(1- \frac{\sigma_B^2}{\sigma_B^2 + \sigma_C^2}(1-2^{-2R_h})\right)}{D}\right)\right]^+, \\
\Delta &\ge \max\left\{\frac{1}{2}\log \frac{\sigma_W^2 + \sigma_A^2 + \sigma_B^2}{ \sigma_A^2 + \sigma_B^2}, \right.\\
& \left.\qquad\qquad \frac{1}{2}\log \frac{\sigma_W^2 + \sigma_A^2 + \sigma_B^2}{\sigma_A^2 + \sigma_B^2} + \frac{1}{2}\log \frac{\sigma_B^2}{2^{2R_h}D} \right\}.
\end{align*}
\end{proposition}
\begin{IEEEproof}
We begin with the converse. For any sequence of $(n, 2^{nR}, 2^{nR_h})$ code that achieves distortion $D$, the minimum rate required in the absence of any information leakage constraint is lower bounded by~\cite[Corollary 12]{Permuter--Steinberg--Weissman2010}
\begin{align}
R \ge \frac{1}{2}\log\left(\frac{\sigma_{B}^2\left(1- \frac{\sigma_B^2}{\sigma_B^2 + \sigma_C^2}(1-2^{-2R_h})\right)}{D}\right). \label{qg_eqn1}
\end{align}
On the other hand, consider now a sequence of $(n, 2^{nR}, 2^{nR_h})$ codes that achieves $(D, \Delta)$. For an $(n, 2^{nR}, 2^{nR_h})$ code that achieves $(D + \e_n, \Delta + \e_n)$, we have the straightforward bound of{\allowdisplaybreaks
\begin{align*}
\Delta + \e_n &\ge I(X^n;W^n) \\
& = \sum_{i=1}^n I(X_i;W_i) \\
& = \frac{n}{2}\log \frac{\sigma_W^2 + \sigma_A^2 + \sigma_B^2}{ \sigma_A^2 + \sigma_B^2}.
\end{align*}}
We also have, following the same arguments as in the converse proof for Proposition~\ref{prop6} (see inequality~\eqref{ineq1}),{\allowdisplaybreaks
\begin{align*}
 n\Delta + nR_h + n\e_n & \ge I(X^n;W^n) + I(X^n;M, M_h|Z^n).
\end{align*}
We now further lower bound this term by
\begin{align*}
 n\Delta + nR_h + n\e  &\stackrel{(a)}{\ge} I(X^n;W^n)+ \sum_{i=1}^n I(X_i; M, M_h, Z_{i+1}^n, Z^{i-1}, X^{i-1}|Z_i) \\
& \stackrel{(b)}{=}  I(X^n;W^n)+\sum_{i=1}^n I(X_i; M, M_h, Z_{i+1}^n, Z^{i-1}, \Xh_i|Z_i) \\
& \ge I(X^n;W^n)+\sum_{i=1}^n I(X_i; \Xh_i|Z_i) \\
& \stackrel{(c)}{=}  nI(X;W)+ n I(X; \Xh|Z, Q) \\
&\ge  nI(X;W)+n I(X; \Xh|Z) \\
& \ge nI(X;W)+ n h(X|Z) -n h(X - \Xh) \\
& \ge nI(X;W)+ n \frac{1}{2}\log \frac{\sigma_B^2}{D+ \e_n}.
\end{align*}}
$(a)$ follows from the i.i.d. property of the $X^n$ and $Z^n$; $(b)$ follows from $\Xh_i$ being a function of $Z^n, M, M_h$; and $(c)$ follows from defining $Q\sim \U[1:n]$,$X_Q = X$, $Z_Q = Z$, $\Xh_Q = \Xh$, $Y_Q = Y$ and $W_Q = W$. The final step follows from the distortion constraint: $\E \sum_{i=1}^n (X_i - \Xh_i)^2/n = \E (X-\Xh)^2 \le D$. Hence, $h(X- \Xh) \le \frac{1}{2}\log 2 \pi e D+ \e_n$. Finally, since $\e_n \to 0$ as $n \to \infty$, we obtain the following bound on $\Delta$. 
\begin{align}
\Delta &\ge \max\left\{\frac{1}{2}\log \frac{\sigma_W^2 + \sigma_A^2 + \sigma_B^2}{ \sigma_A^2 + \sigma_B^2}, \frac{1}{2}\log \frac{\sigma_W^2 + \sigma_A^2 + \sigma_B^2}{\sigma_A^2 + \sigma_B^2} + \frac{1}{2}\log \frac{\sigma_B^2}{2^{2R_h}D} \right\}. \label{qg_eqn2}
\end{align}

We now turn to the achievability proof for the lower bounds for $R$ and $\Delta$ in inequalities~\eqref{qg_eqn1} and~\eqref{qg_eqn2}, respectively. We use Proposition~\ref{prop4} and set $U = \emptyset$, $U_h =Y + N_h$ and $V = X + N_e$, where $N_h\sim N(0, \sigma_h^2)$ and $N_e\sim N(0, \sigma_e^2)$ are independent Gaussian random variables. These definitions result in the Markov Chain $V-X-(U_h, Y,Z,W)$. We set $\Xh = \E(X|U_h, V, Z)$. It suffices to consider only the case of $D \le \sigma_{B}^2\left(1- \frac{\sigma_B^2}{\sigma_B^2 + \sigma_C^2}(1-2^{-2R_h})\right)$. Let 
\begin{align*}
\sigma_h^2 &= \frac{\sigma_B^2 + \sigma_C^2}{2^{2R_h} - 1}, \\
\sigma_{X|U_h, Z}^2 &= \sigma_{B}^2\left(1- \frac{\sigma_B^2}{\sigma_B^2 + \sigma_C^2}(1-2^{-2R_h})\right),\\
\sigma_e^2 &= \frac{\sigma_{X|U_h,Z}^2 D}{\sigma_{X|U_h,Z}^2 - D}
\end{align*}
With these definitions, we have the following quantities.{\allowdisplaybreaks
\begin{align*}
\Var(X|U_h, Z, V) &= \E (X - \E(X|U_h, Z, V))^2 \\
&= \E (B - \E(B|B+C+N_h, B+ N_e))^2 \\
& = D, \\
\Var(X|U_h, Z) &= \sigma_{X|U_h, Z}^2, \\
I(Y;U_h|Z) & = R_h, \\
h(X|U_h, Z) & = \frac{1}{2}\log 2\pi e \sigma_{X|U_h, Z}^2, \\
h(X|U_h, V,Z) & = \frac{1}{2}\log 2\pi e D.
\end{align*} }
It is now straightforward to verify that the achievability scheme in Proposition~\ref{prop4} achieves the outer bound with these choice of auxiliary random variables, which completes the proof. 
\end{IEEEproof}
Similarly, in a close analog to Corollary~\ref{coro8}, we have the following R.D.I. characterization for another Quadratic Gaussian setting.
\begin{proposition} \label{qg2}
For the setting in Figure~\ref{fig2}, let $X\sim N(0, \sigma_X^2)$, $Z = X+ A$, $W = Z+B$, $Y = W + C$, and $A\sim N(0, \sigma_A^2)$, $B\sim N(0, \sigma_B^2)$ and $C\sim N(0, \sigma_C^2)$ be mutually independent Gaussian random variables, and the distortion measure be squared loss.  To avoid degenerate cases, we assume that $\sigma^2_X, \sigma^2_A, \sigma^2_B, \sigma^2_C > 0$. Then, the R.D.I. region is given by {\allowdisplaybreaks
\begin{align*}
R_h &\ge 0, \\
R & \ge \left[\frac{1}{2}\log\left(\frac{\sigma_X^2\sigma_A^2}{(\sigma_X^2 + \sigma_A^2)D}\right)\right]^+, \\
\Delta &\ge \max\{\frac{1}{2}\log\left(\frac{\sigma_X^2+\sigma_A^2+ \sigma_B^2}{\sigma_A^2+ \sigma_B^2}\right), \frac{1}{2}\log\left(\frac{\sigma_X^2+\sigma_A^2+ \sigma_B^2}{\sigma_A^2+ \sigma_B^2}\right) + \frac{1}{2}\log\left(\frac{\sigma_X^2\sigma_A^2}{(\sigma_X^2 + \sigma_A^2)D}\right) - R_h\}.
\end{align*}  }
\end{proposition}
Proof of this Proposition is given in Appendix~\ref{appen:d}.
\section{Amplification Measures} \label{sect:6} 
We now turn our attention to source amplification measures at the decoder. Instead of symbol by symbol distortion measures like those considered in the previous sections, we consider the following two amplification measures. Let $U^n_{\rm dec}$ be the overall information at the decoder, which includes the decoder's S.I. and the message(s) received.
\begin{itemize}
\item List constraint: Based on the decoder's information, it forms a list, $\mathcal{L}(U^n_{\rm dec})$, of $x^n$ sequences such that $|\mathcal{L}(U^n_{\rm dec})| \le 2^{nD}$ and $\P(X^n \in \mathcal{L}(U^n_{\rm dec})) \to 1$ as $n \to \infty$. The list constraint is a straightforward generalization of lossless source coding, with $D = 0$ corresponding to the lossless case.
\item Entropy constraint: Here, we wish to ensure that $\limsup_{n\to \infty}\frac{1}{n}H(X^n|U^n_{\rm dec}) \le D$. The entropy constraint can be shown to be equivalent to \textit{block log-loss} constraint~\cite{Coutrade2012a}. That is, the decoder's reconstruction vector is the set of all probability distributions over $|\Xc|^n$, and the distortion is measured by $\log(1/\xh(x^n))/n$, where $\xh(x^n)$ is the estimated probability of $X^n = x^n$. Block log-loss is a strengthening of the symbol-by-symbol log-loss distortion measure defined in Corollary~\ref{coro2} since it allows more general probability distributions over $|\Xc|^n$ instead of only product distributions (in the case of symbol by symbol log loss).   
\end{itemize}

We now consider how the R.D.I. regions change when we replace log-loss distortion constraint with the amplification measures.
\begin{proposition} \label{prop5}
For the settings in Corollaries~\ref{coro2},~\ref{coro5} and~\ref{coro8}, and Proposition~\ref{prop6}, the R.D.I. regions remain unchanged if the log-loss distortion measure at the decoder is replaced by a list or entropy constraint.
\end{proposition}

For the case of entropy constraint (or block log-loss), Proposition~\ref{prop5} states that even if we allow more general probability distributions than the product distributions for symbol-by-symbol log-loss, there is no gain in the R.D.I. regions for our settings. In the case of list constraint, it relates achievable distortion under log-loss to the exponent of the achievable list size, and also provides a way of recovering results for lossless source coding from results for log-loss distortion measure with $D$ set to zero.
\begin{IEEEproof}

In our proof, we will use the following lemma found in~\cite{Kim--Sutivong--Cover2008}, adapted to our notation. 
\begin{lemma}~\label{lemlist}
Let $\mathcal{L}(U_{\rm dec}^n)$ be a sequence of list decoders such that $\P(X^n \notin \mathcal{L}(U_{\rm dec}^n)) \to 0$ as $n \to \infty$. Then,
\begin{align*}
H(X^n|U^n_{\rm dec}) \le \log |\mathcal{L}(U_{\rm dec}^n)| + n \e_n,
\end{align*}
where $\e_n \to 0$ as $n \to \infty$.
\end{lemma}
\subsection*{Achievability under list decoding}
We now show the achievability of Corollaries~\ref{coro2},~\ref{coro5} and~\ref{coro8}, and Proposition~\ref{prop6}, when the log-loss constraint at the decoder is replaced by a list constraint, with $\log|\mathcal{L}(U^n_{\rm dec})| \le nD$. Let $V_{\rm dec}^n$ denote all the codewords decoded and the original side information at the decoder for Corollary~\ref{coro2} and Propositions~\ref{coro3} and~\ref{coro4}. In the achievability scheme of Corollaries~\ref{coro2},~\ref{coro5} and~\ref{coro8}, and Proposition~\ref{prop6}, recall that our scheme results in $\P((V_{\rm dec}^n,X^n) \in \aep ) \to 1$ as $n \to \infty$. The list decoder forms the following list:
\begin{align*}
\mathcal{L}(v^n_{\rm dec}):= \{x^n: (x^n, v^n_{\rm dec}) \in \aep\}.
\end{align*}
From properties of typical sequences (see~\cite[Chapter 2]{El-Gamal--Kim2010}), we have that
\begin{align*}
\frac{1}{n}\log|\mathcal{L}(v^n_{\rm dec})| &\le H(X|V_{\rm dec}) + \d(\e)\\
& = D + \d(\e).
\end{align*} 
The last step follows from the choice of auxiliary random variables in Corollaries~\ref{coro2},~\ref{coro5} and~\ref{coro8}, and Proposition~\ref{prop6}. The requirement that $\P(X^n \in \mathcal{L}(V^n_{\rm dec})) \to 1$ as $n \to \infty$ follows from $\P((V_{\rm dec}^n,X^n) \in \aep ) \to 1$ as $n \to \infty$ in our achievability scheme. 
\subsection*{Achievability under entropy constraint}
Achievability under entropy constraint is a straightforward consequence of achievability under list constraint and Lemma~\ref{lemlist}. Since we have a sequence of list decoders satisfying the conditions in Lemma~\ref{lemlist},
\begin{align*}
\frac{1}{n}H(X^n|V^n_{\rm dec}) &\le \frac{1}{n}\log |\mathcal{L}(V_{\rm dec}^n)| +  \e_n \\
& \le D+ \d(\e) + \e_n.
\end{align*}
\subsection*{Converse}
From Lemma~\ref{lemlist}, any code under list constraint that achieves a list size of $D_{\rm list}$ is also a code that achieves a block log-loss (or entropy constraint) of at most $D_{\rm list} - \e_n$. Hence, any outer bound for our settings under entropy constraint is also an outer bound for our settings under the list constraint. We therefore only need to consider outer bounds for our settings under the entropy constraint in the converse. 

With the above observation, recall that in our proof of converse for Proposition~\ref{prop6}, a key property of log-loss that we used is the fact that log-loss distortion upper bounds the entropy of the source sequence given the overall side information at the decoder (see Lemma~\ref{lemld}). Similar to log-loss, given a code with entropy constraint of $D_{\rm entropy}$, we have, by definition, the following upper bound on the entropy of the source sequence given the overall side information at the decoder.
\begin{align}
\frac{1}{n} H(X^n|U^n_{\rm dec}) &\le D_{\rm entropy}. \label{ent} 
\end{align}
It can be verified that our converse proof for Proposition~\ref{prop6} continues to hold under the entropy constraint with the upper bound in Lemma~\ref{lemld}, $\frac{1}{n} H(X^n|U^n_{\rm dec}) \le D_{\rm log-loss}$, being replaced by inequality~\eqref{ent}. For Corollaries~\ref{coro2},~\ref{coro5} and~\ref{coro8}, the upper bound $\frac{1}{n} H(X^n|U^n_{\rm dec}) \le D_{\rm log-loss}$ was used implicitly in the proofs of converse, and similarly, it can be verified that the proof of converse continues to hold with inequality~\eqref{ent} for the entropy constraint case. The details are given in Appendix~\ref{appen:amp}. 
\end{IEEEproof}
\begin{remark}
The property $\Rc_{\rm SI-Enc}(\Yt) = \Rc_{\rm WZ}(\Yt)$ enjoyed by the log-loss distortion measure was used to obtain the R.D.I. regions under log-loss for Corollaries~\ref{coro2},~\ref{coro5} and~\ref{coro8}. Using inequality~\eqref{ent} and Lemma~\ref{lemld}, we can show that the same property also holds true under block log-loss or list constraint. This property can also be used to give proofs of converse for Corollaries~\ref{coro2},~\ref{coro5} and~\ref{coro8}, similar to what was done in the log-loss case. 
\end{remark}

\section{Conclusion}\label{sect:7}
We considered the setting of secure lossy source coding when either coded or uncoded S.I. is available at the decoder. For the case of uncoded side information, we considered two related settings. Our first setting considered the case where the eavesdropper's S.I. is not available at the decoder. We gave general inner and outer bounds for this setup, and characterized the R.D.I. region for some special cases. We then considered the second uncoded S.I. setting where the eavesdropper's S.I. is also available to the decoder. For this case, we again give general inner and outer bounds for this setting and characterized the R.D.I. region for some special cases. The main idea used in the achievability proofs for these settings is in the generation of a secret key, via binning the S.I. at the encoder and the decoder, to reduce the information leakage rate at the eavesdropper. This idea can also be used in other secure source coding settings~\cite{kk2013}. A recurring theme in the special cases for which we were able to find the R.D.I. regions is that the source, S.I.s and distortion measure satisfy the condition that S.I. at the encoder does not improve the rate-distortion region.

We then considered the case of coded S.I. at the encoder and decoder. For this case, we gave an achievability scheme for the general setting that used the idea of generating a secret key from the coded S.I., as well as the helper generating an independent secret key for both the encoder and the decoder. We characterized the R.D.I. regions for several settings and recovered previous results in the literature as special cases of our settings. Finally, we considered two amplification measures for the decoder, list-decoding and entropy minimization, and showed that the R.D.I. regions under these measures coincide with the R.D.I. region under per symbol log-loss for the cases we considered in this paper.  
 
\section*{Acknowledgment}
We thank Prof. Tsachy Weissman of Stanford University, Profs. Mikael Skoglund and Tobias Oechtering of KTH Sweden for helpful discussions. 

\bibliographystyle{IEEEtran}
\bibliography{sienc}
\appendices
\section{Proof of Proposition~\ref{prop1}} \label{appen:prop1}
Given a $(n, 2^{nR})$ code that achieves $(D+ \e_n, \Delta+ \e_n)$, define the auxiliary random variables $U_i :=(M, Y^{i-1}, Z_{i+1}^n)$ and $V_i = (Y_{i+1}^n, X_{i+1}^n)$ for $i \in [1:n]$. A lower bound on the rate is then given by{\allowdisplaybreaks
\begin{align*}
nR &\ge H(M) \\
&\ge I(X^n;M|Y^n) \\
& = \sum_{i=1}^n I(X_i; M|Y^n, X_{i+1}^n) \\
& \stackrel{(a)}{=}\sum_{i=1}^n I( X_i; M, X_{i+1}^n, Y_{i+1}^n, Y^{i-1}|Y_i ) \\
& \stackrel{(b)}{=}\sum_{i=1}^n I( X_i; M, X_{i+1}^n, Y_{i+1}^n, Y^{i-1}, Z_{i+1}^n|Y_i ) \\
& = \sum_{i=1}^n I( X_i; U_i, V_i|Y_i ).
\end{align*}}
The last step follows from the definition of $U_i$and $V_i$. $(a)$ follows from $(X^n, Y^n)$ being generated i.i.d. and $(b)$ follows from the Markov Chain $Z_{i+1}^n - ( M, X_{i+1}^n, Y_{i+1}^n, Y^{i}) - X_i$.
For the information leakage term, we have {\allowdisplaybreaks
\begin{align*}
 n\Delta + \e_n &= I(X^n; M, Z^n) \\
& = I(X^n, Y^n; M, Z^n) - I(Y^n;M, Z^n|X^n) \\
& = I(X^n, Y^n;Z^n) + I(X^n, Y^n;M|Z^n)  - I(Y^n;M, Z^n|X^n) \\
& = I(X^n, Y^n;Z^n) + I(X^n, Y^n;M) - I(M;Z^n)  - I(Y^n;M, Z^n|X^n) \\
& = \sum_{i=1}^n I(X_i, Y_i;Z_i) + I(X^n;M|Y^n) + I(M;Y^n) - I(M;Z^n) - I(Y^n;M, Z^n|X^n) \\
& \stackrel{(a)}{=} \sum_{i=1}^n I(X_i, Y_i;Z_i) + I(X^n;M|Y^n) + \sum_{i=1}^n (I(M, Y^{i-1}, Z_{i+1}^n; Y_i) - I(M, Y^{i-1}, Z_{i+1}^n; Z_i)) \\
& \qquad - I(Y^n;M, Z^n|X^n) \\
& = \sum_{i=1}^n I(X_i, Y_i;Z_i)+ I(X^n;M|Y^n) + \sum_{i=1}^n (I(U_i; Y_i) - I(U_i; Z_i)) - I(Y^n;M, Z^n|X^n) \\
& = \sum_{i=1}^n I(X_i, Y_i;Z_i) + \sum_{i=1}^nI(X_i;M|Y^n, X_{i+1}^n) + \sum_{i=1}^n (I(U_i; Y_i) - I(U_i; Z_i))\\
& \qquad - I(Y^n;M, Z^n|X^n) \\
& = \sum_{i=1}^n I(X_i, Y_i;Z_i) + \sum_{i=1}^nI(X_i;M, Y^{i-1}, Y_{i+1}^n, X_{i+1}^n|Y_i) + \sum_{i=1}^n (I(U_i; Y_i) - I(U_i; Z_i)) \\
& \qquad - I(Y^n;M, Z^n|X^n) \\
& = \sum_{i=1}^n I(X_i, Y_i;Z_i) + \sum_{i=1}^nI(X_i;M, Y^{i-1}, Y_{i+1}^n, X_{i+1}^n, Z_{i+1}^n|Y_i) \\
& \qquad + \sum_{i=1}^n (I(U_i; Y_i) - I(U_i; Z_i)) - I(Y^n;M, Z^n|X^n).
\end{align*}}
$(a)$ follows from the Csisz\'{a}r Sum lemma. The lower bound
\begin{align*}
\Delta + \e_n \ge \sum_{i=1}^n I(X_i;Z_i)
\end{align*}
is straightforward to show.

Now, let $Q \sim \U[1:n]$ be the time-sharing random variable that is independent of all other random variables. Define $U = (Q, M, Y^{Q-1}, Z_{Q+1}^n)$, $V = (Y_{Q+1}^n, X_{Q+1}^n)$ and $(X_Q, Y_Q, Z_Q) = (X,Y,Z)$. Then,{\allowdisplaybreaks
\begin{align*}
R &\ge \frac{1}{n}\sum_{i=1}^n I( X_i; U_i, V_i|Y_i, Q = i ) \\
& = I( X; U_Q, V|Y, Q) \\
& = I(X;U,V|Y).
\end{align*}}
The last step follows from the fact that $(X^n, Y^n)$ is i.i.d. and hence, $I(X;Q|Y) = 0$. Next,{\allowdisplaybreaks
\begin{align*}
\Delta + \e_n &\ge I(X,Y;Z|Q) + I(X;U_Q,V|Y, Q) + I(U_Q;Y|Q) - nI(U_Q;Z|Q) - \frac{1}{n}I(Y^n;M, Z^n|X^n)\\
& = I(X,Y;Z) + I(X;U,V|Y) + I(U;Y) - I(U;Z) - \frac{1}{n}I(Y^n;M, Z^n|X^n) \\
& = I(X,Y;Z) + I(X, Y;U,V) - I(Y;U,V) + I(U;Y) - I(U;Z) - \frac{1}{n}I(Y^n;M, Z^n|X^n) \\
& = I(X,Y;Z) + I(X, Y;U,V|Z) +I(V;Z|U) - I(V;Y|U) - \frac{1}{n}I(Y^n;M, Z^n|X^n) \\
& = I(X,Y;U,V, Z) +I(V;Z|U) - I(V;Y|U) - \frac{1}{n}I(Y^n;M, Z^n|X^n) \\
& \ge I(X;U,V,Z)+ I(V;Z|U) - I(V;Y|U)+ I(Y;U,V,Z|X) - \frac{1}{n}H(Y^n|X^n) \\
& = I(X;U,V,Z)+ I(V;Z|U) - I(V;Y|U) - H(Y|U,V,X,Z). 
\end{align*}
Finally, we consider the bound on distortion. We have
\begin{align*}
D + \e_n &\ge \frac{1}{n}\sum_{i=1}^n \E d(X_i, \xh_i(Y^n, M)) \\
& \ge \frac{1}{n}\sum_{i=1}^n \E d(X_i, \xh'_i(U_i, V_i, Y_i)) \\
& = \E_Q \E(d(X_Q, \xh'_Q(U_Q, V_Q, Y_Q))|Q) \\
& = \E d(X, \xh'(U,V,Y)).  
\end{align*}}
Hence, the choice of auxiliary random variables satisfy the distortion constraint with reconstruction function $\xh'$. Next, noting that $\e_n \to 0$ as $n \to \infty$ then gives us the required bound. The Markov Chain condition $(U,V)-(X,Y) - Z$ follows from the definition of the auxiliary random variables and is straightforward to verify.

It remains to give upper bounds on the cardinalities of $U$ and $V$. The stated bounds follow straightforwardly from the cardinality bounding techniques in~\cite[Appendix C]{El-Gamal--Kim2010} and we omit them here. 

\section{Proof of Proposition~\ref{prop2}} \label{appen:prop2}
We give a proof of the lower bound, with details for the fairly standard decoding steps left out of the proof.  In our proof, we will use the following lemma.
\begin{lemma} \label{lem2}
Fix $\e>0$. Let $Y^n \sim \prod_{i=1}^n p(y_i)$ and let $W^n$ be a random variable such that \\$\P((Y^n, W^n) \in \aep(Y,W)) \to 1$ as $n \to \infty$. Bin the set of all $|\Yc|^n$ sequences to $2^{nR_K}$ bins uniformly at random, and let $K$ be the bin index such that $Y^n \in \Bc(K)$. Then, if $R_K \le H(Y|W)$,
\begin{align*}
H(Y^n|W^n, K) \le n H(Y|W) -n R_K + n \d(\e).
\end{align*}
\end{lemma}
Proof of this lemma is given in Appendix~\ref{appen:1}. We note here that the special case of $R_K = 0$ will be used several times in the proofs of this proposition and Proposition~\ref{prop4}.
\subsection*{Codebook generation}
We generate two codebooks, the rate-distortion codebook and the key generation codebook. We first start with the rate distortion codebook, $\Cc_{\rm RD}$. 
\begin{itemize}
\item Generate $2^{n(I(U;X,Y) + \d(\e))}$ $U^n(l_0)$ sequences according to $\prod_{i=1}^n p(u_i)$, $l_0 \in [1: 2^{n(I(U;X,Y) + \d(\e))}]$.
\item For each $u^n(l_0)$ sequence, generate $2^{n(I(V;X,Y|U) + \d(\e))}$ $V^n(l_1, l_0)$ sequences according to \\$\prod_{i=1}^n p(v_i|u_i)$, $l_1 \in [1: 2^{n(I(V;X,Y|U) + \d(\e))}]$.  
\item Partition the set of $U^n$ sequences to $2^{n(I(U;X|Y) + 3\d(\e))}$ bins, $\Bc_{\rm RD}(m_0)$, $m_0 \in [1: 2^{n(I(U;X|Y) + 3\d(\e))}]$.
\item For each $l_0$, partition the set of $V^n$ sequences to $2^{n(I(V;X|Y,U) + 3\d(\e))}$ bins, $\Bc_{\rm RD}(m_1,l_0)$,\\ $m_1 \in [1: 2^{n(I(V;X|Y,U) + 3\d(\e))}]$.
\end{itemize}
This completes the codebook generation for $\Cc_{\rm RD}$. We now turn to the key generation codebook, $\Cc_{\rm K}$, which has only a single step. We assume that $\min\{I(V;X|Y,U), H(Y|X,Z,U,V)\}>0$. Otherwise, no binning is done.
\begin{itemize}
\item Randomly and uniformly bin the set of $Y^n$ sequences to $2^{nR_K}$ bins, $\Bc_{\rm K}(m_k)$, where \\$R_{K}:= \min\{H(Y|U,V,X,Z), I(V;X|U,Y)\}$ and $m_k \in [1:2^{nR_K}]$.
\end{itemize}
We use $\Cc:=\{\Cc_{\rm RD}, \Cc_{\rm K}\}$ to denote the combined codebook.
\subsection*{Encoding}
\begin{itemize}
\item Given sequences $(x^n, y^n)$, the encoder first looks for a sequence $u^n(l_0)$ such that $(u^n(l_0),x^n, y^n)\in \aep$. If there is more than one such sequence, the encoder selects one sequence uniformly at randomly from the set of jointly typical $u^n$ sequences. If there is none, the encoder randomly and uniformly selects a sequence $u^n$ from the set of all sequences.
\item Next, the encdoer looks for a $v^n(l_1,l_0)$ such that $(v^n(l_1,l_0),u^n(l_0),x^n, y^n)\in \aep$. If there is more than one such sequence, the encoder selects one sequence uniformly at random from the set of jointly typical $v^n$ sequences. If there is none, the encoder randomly and uniformly selects a sequence $v^n$ from the set of all sequences.
\item The encoder then looks for the index $m_0$ and $m_1$ such that $u^n(l_0) \in \Bc_{\rm RD}(m_0)$ and $v^n(l_1,l_0) \in \Bc_{\rm RD}(m_1, l_0)$. 
\item Next, it splits the index $m_1$ into two parts, $m_{1s} \in [1:2^{nR_K}]$ and $m_{1o} \in[1:2^{n(I(V;X|U,Y) + 3\d(\e) - R_K)}]$.
\item The encoder then looks for the index $m_k$ such that $y^n \in \Bc_{\rm K}(m_k)$.
\item Finally, the encoder sends out the indices $m_0$, $m_{1o}$ and $m_{1s}\oplus m_{k}$\footnote{Here, $m_{1s}\oplus m_{k}$ denotes the modulo operation, $(m_{1s} + m_{k}){\rm mod}\, 2^{nR_K}$, with the exception that $0$ is mapped to $2^{nR_K}$.}, resulting in a rate of \\$I(X;U,V|Y) + 6\d(\e)$. 
\end{itemize}
\subsection*{Analysis of distortion}
Since the decoder has the sequence $y^n$, it first finds $m_k$ to unscramble $m_{1s}\oplus m_{k}$, thereby recovering the index $m_1$. It then decodes the codewords $u^n(L_0)$ and $v^n(L_0, L_1)$ using successive decoding. That is, it first looks for a $\lh_0$ such that $(u^n(\lh_0), y^n) \in \aep$ and $u^n(\lh_0)\in \Bc(m_0)$. An error occurs if there is no such $\lh_0$. Next, it then looks for a $\lh_1$ such that $(v^n(\lh_1, \lh_0), u^n(\lh_0), y^n) \in \aep$ and $v^n(\lh_0, \lh_1)\in \Bc(m_1, \lh_0)$. Similarly, an error occurs if there is no such $\lh_1$. The analysis of the probability of error follows quite straightforwardly from the analysis for the Wyner-Ziv setting in~\cite[Chapter 11]{El-Gamal--Kim2010}, and we will omit it here. From the rates given in the codebook generation and encoding process, it can be shown that the probability of error ($\lh_0 \neq L_0$ or $\lh_1 \neq L_1$), averaged over codebooks, goes to zero as $n \to \infty$. 

Further, from the rates given and the covering lemma in~\cite[Chapter 3]{El-Gamal--Kim2010}, we have that \\$\P((U^n(L_0), V^n(L_0, L_1), X^n, Y^n) \in \aep) \to 1$ as $n \to \infty$. Hence, following~\cite[Chapter 3]{El-Gamal--Kim2010}, the expected distortion, averaged over codebooks, is less than or equal to $D + \d(\e)$ as $n \to \infty$. 

\subsection*{Analysis of information leakage rate}
For notational convenience, we will use $\d(\e)$ to denote all terms that go to zero as $\e \to 0$, or $n \to \infty$.
\begin{align}
n\Delta & = I(X^n; Z^n, M_0, M_{1o}, M_{1s}\oplus M_K| \Cc)  \nonumber\\
& \le I(X^n; Z^n, L_0, M_{1o}, M_{1s}\oplus M_K| \Cc)  \nonumber\\
& = I(X^n, Y^n; Z^n, L_0, M_{1o}, M_{1s}\oplus M_K|\Cc) - I(Y^n; Z^n, L_0, M_{1o}, M_{1s}\oplus M_K|X^n, \Cc). \label{ie0}
\end{align}
We now bound each of the terms separately.{\allowdisplaybreaks
\begin{align}
& I(X^n, Y^n; Z^n, L_0, M_{1o}, M_{1s}\oplus M_K|\Cc)\nonumber\\
& = H(Z^n, L_0|\Cc) + H(M_{1o}, M_{1s} \oplus M_K|L_0, Z^n, \Cc) -  H(Z^n, L_0, M_{1o}, M_{1s}\oplus M_K|X^n, Y^n, \Cc) \nonumber\\
& \le H(Z^n, L_0|\Cc) + H(M_{1o}, M_{1s} \oplus M_K|L_0,\Cc) -  H(Z^n|X^n, Y^n, \Cc) \nonumber\\
& \le H(L_0|\Cc) + H(Z^n|L_0, \Cc) + n I(V;X|U,Y)  - nH(Z|X,Y) + n\d(\e)\nonumber\\
& \le H(L_0|\Cc) + H(Z^n|U^n(L_0)) + n I(V;X|U,Y) - nH(Z|X,Y)+ n\d(\e) \nonumber\\
& \stackrel{(a)}{\le} nI(U;X,Y) + n\e + nH(Z|U) + n I(V;X|U,Y) - nH(Z|X,Y)+ n\d(\e) \nonumber\\
& = nI(X,Y;Z,U) + nI(V;X|U,Y) + n\d(\e). \label{ie1}
\end{align}}
The final step uses the Markov relation $U-(X,Y) - Z$. In $(a)$, we applied Lemma~\ref{lem2} to $H(Z^n|U^n(L_0))$. The condition that $\P((U^n(L_0), Z^n) \in \aep) \to 1$ as $n \to \infty$ follows from the rates given, the codebook generation and encoding process, and the conditional typicality lemma and covering lemma in~\cite{El-Gamal--Kim2010}. For the second term, we have{\allowdisplaybreaks
\begin{align}
& - I(Y^n; Z^n, L_0, M_{1o}, M_{1s}\oplus M_K|X^n, \Cc) \nonumber\\
& = -H(Y^n|X^n,  \Cc) + H(Y^n|X^n, Z^n, L_0, M_{1o}, M_{1s}\oplus M_K, \Cc) \nonumber\\
& \le -nH(Y|X) + H(Y^n, L_1|X^n, Z^n, L_0, M_{1o}, M_{1s}\oplus M_K, \Cc) \nonumber\\
& = -nH(Y|X) + H(L_1|X^n, Z^n, L_0, M_{1o}, M_{1s}\oplus M_K, \Cc)+ H(Y^n|X^n, Z^n, L_0, L_1, M_k, \Cc) \nonumber\\
& \le -nH(Y|X) + H(L_1|X^n, Z^n, L_0, M_{1o}, M_{1s}\oplus M_K, \Cc)+ H(Y^n|X^n, Z^n,U^n(L_0), V^n(L_0,L_1), M_k) \nonumber\\
& \stackrel{(a)}{\le}  -nH(Y|X) + H(L_1|X^n, Z^n, L_0, M_{1o}, M_{1s}\oplus M_K, \Cc)+ n H(Y|U,V,X,Z) - nR_K + n\d(\e)\nonumber\\
& \le -nH(Y|X) + H(L_1|X^n, Z^n, L_0, \Cc)+ n H(Y|U,V,X,Z) - nR_K + n\d(\e)\nonumber\\
& \le -nH(Y|X) + nI(V;Y|U,X,Z)+ n H(Y|U,V,X,Z) - nR_K + n\d(\e).  \label{ie2}
\end{align}}
In $(a)$, we apply Lemma~\ref{lem2} to $H(Y^n|X^n, Z^n,U^n(L_0), V^n(L_0,L_1), M_k)$. To check that the conditions for applying Lemma~\ref{lem2} are satisfied, observe that $R_K  = \min\{I(V;X|Y,U), H(Y|X,Z,U,V)\}  \le  H(Y|X,Z,U,V)$. The condition that \\$\P((U^n(L_0), V^n(L_0,L_1), X^n, Y^n,Z^n)\in \aep) \to 1$ follows again from the rates given and the encoding process. In the final step, we upper bound $H(L_1|X^n, Z^n, L_0, \Cc)$ as follow.{\allowdisplaybreaks 
\begin{align*}
& H(L_1|X^n, Z^n, L_0, \Cc) \\
& = H(L_1, X^n, Z^n| L_0, \Cc) - H(X^n, Z^n|L_0, \Cc) \\
& = H(L_1| L_0, \Cc)+ H(X^n, Z^n| L_0, L_1, \Cc) - H(X^n, Z^n| \Cc)+ H(L_0|\Cc) - H(L_0|X^n, Z^n, \Cc) \\
& = H(L_1| L_0, \Cc)+ H(X^n, Z^n| L_0, L_1, \Cc) - H(X^n, Z^n| \Cc)+ H(L_0|\Cc) - H(L_0|X^n, Y^n, Z^n, \Cc) \\
& \qquad - I(Y^n;L_0|X^n, Z^n, \Cc)\\
& \le nI(V;X,Y|U)+ H(X^n, Z^n| U^n(L_0), V^n(L_0,L_1)) - nH(X,Z)+ nI(U;X,Y) \\
&\qquad - I(Y^n;L_0|X^n, Z^n, \Cc) + n\d(\e)\\
& \stackrel{(a)}{\le} nI(V;X,Y|U)+ H(X,Z|U,V) - nH(X,Z)+ nI(U;X,Y) - I(Y^n;L_0|X^n, Z^n, \Cc) + n\d(\e)\\
&  \le nI(V;X,Y|U)+ H(X,Z|U,V) - nH(X,Z)+ nI(U;X,Y) - H(Y^n|X^n, Z^n, \Cc) \\
& \qquad+ H(Y^n|U^n(L_0), X^n, Z^n)  + n\d(\e)\\
& \stackrel{(b)}{\le} nI(V;X,Y|U)+ H(X,Z|U,V) - nH(X,Z)+ nI(U;X,Y) - nH(Y|X,Z) + nH(Y|X,Z,U)  \\
& \qquad + n\d(\e)\\
& =  nI(V;Y|U,X,Z) + n\d(\e). 
\end{align*}}
$(a)$ and $(b)$ follow from applying Lemma~\ref{lem2} to the terms $H(X^n, Z^n| U^n(L_0), V^n(L_0,L_1))$ and \\$H(Y^n|U^n(L_0), X^n, Z^n)$ respectively. The final step uses the Markov condition $(V,U) - (X,Y) - Z$ and hence, $I(V,U;X,Y) = I(V,U;X,Y,Z)$.

Combining the bounds in~\eqref{ie1} and~\eqref{ie2} into~\eqref{ie0} then leads us to{\allowdisplaybreaks 
\begin{align*}
\Delta &\le  I(X,Y;Z, U) + I(V;X|U, Y) - H(Y|X) + H(Y|X,Z, V,U) - R_K + H(Y|U,X,Z) \\
& \qquad - H(Y|X,Z,V, U) -\d(\e)\\
& = I(X;Z,U) + I(V;X|U,Y) - R_K.
\end{align*}}
Hence, any $\Delta' > \Delta$ is achievable.

\section{Proof of lemma~\ref{lem2} }\label{appen:1}
Let $\e'' > \e' > \e$ and define $N(w^n, k):= |\{y^n: y^n \in \Bc(k), (y^n,w^n) \in \mathcal{T}_{\e''}^{(n)}\}|$ and $E_1 = 1$ if $N(W^n, K) > a$ and $0$ otherwise. Let $E_2 = 1$ if $(W^n,Y^n) \notin \aep$ and $0$ otherwise. Observe that by assumption, $\P(E_2 =1) \to 0$ as $n \to \infty$. We now focus on $E_1$.{\allowdisplaybreaks   
\begin{align}
&\P(E_1 =1) \nonumber\\
&\le \sum_{w^n \in \aepvar, k}p(w^n, k)\P(E_1=1|W^n = w^n, K= k) + \P(W^n \notin \aepvar) \nonumber\\
& \le \sum_{w^n \in \aepvar, k}p(w^n, k)\P(E_1=1|W^n = w^n, K= k) + \e_n \nonumber\\
& = \sum_{w^n\in \aepvar, k}p(w^n, k)\P(N(w^n, k)> a|W^n = w^n, K= k) + \e_n \nonumber\\
& = \sum_{w^n \in\aepvar, k}p(w^n, k)\sum_{\bar{y}^n}\P(Y^n =\bar{y}^n|W^n = w^n, K= k)\P(N(w^n, k)> a|Y^n = \bar{y}^n, W^n = w^n, K= k) \nonumber\\
& \qquad + \e_n \nonumber\\
& = \sum_{w^n \in\aepvar, k}p(w^n, k)\sum_{\bar{y}^n}\P(Y^n =\bar{y}^n|W^n = w^n, K= k)\P(N(w^n, k)> a|Y^n = \bar{y}^n, K= k) + \e_n. \label{meq1}
\end{align}}
The last line follows from the Markov relation $(W^n = w^n) - (Y^n = \bar{y}^n, K=k) - \{N(w^n, k)>a\}$, which follows from the binning of all $|\Yc|^n$ sequences being done uniformly at random, independent of $W^n$ and $Y^n$. 
{\allowdisplaybreaks
\begin{align}
\P(N(w^n, k)> a|Y^n = \bar{y}^n, K= k) & = \frac{\P(K = k, N(w^n, k)> a|Y^n = \bar{y}^n)}{\P(K=k|Y^n = \bar{y}^n)} \nonumber \\
& \stackrel{(a)}{=} 2^{nR_K}\P(K = k, N(w^n, k)> a|Y^n = \bar{y}^n) \nonumber \\
& \stackrel{(b)}{=} 2^{nR_K} \P(\bar{y}^n\in \Bc(k), N(w^n, k)> a)\nonumber \\
& \stackrel{(c)}{\le}  2^{nR_K} 2^{-nR_K}\mathbf{.}\nonumber\\
& \quad \P(|\{y^n: y^n \in \Bc(k), y^n \neq \bar{y}^n,(y^n,w^n) \in \mathcal{T}_{\e''}^{(n)}\}|>a-1) \nonumber \\
& {\le} \P( N(w^n, k)> a-1). \nonumber 
\end{align} }
$(a)$ follows from $\P(K=k|Y^n = \bar{y}^n) = 2^{nR_K}$. $(b)$ and $(c)$ follow from the fact that the sequences are binned uniformly at random, independent of other sequences. Observe now that $\E  N(w^n, k) = |\mathcal{T}_{\e''}^{(n)}(Y|w^n)|2^{-nR_K}$ since the sequences are binned uniformly at random. Using the bound $|\mathcal{T}_{\e''}^{(n)}(Y|w^n)|\le 2^{n(H(Y|W)+ \d_1(\e''))}$ and applying Markov's inequality with $a-1 = 2^{n(H(Y|W)-R_K + 2\d_1(\e''))}$ to $\P( N(w^n, k)> a-1)$, we have 
\begin{align}
\P(N(w^n, k)> a|Y^n = \bar{y}^n, K= k) \le \frac{1}{2^{n\d_1(\e'')}}. \label{meq2}
\end{align}

Using the bound~\eqref{meq2} in~\eqref{meq1}, we obtain
\begin{align*}
\P(E_1 =1) \le \frac{1}{2^{n\d_1(\e'')}} + \e_n.
\end{align*}
Hence, with $a = 2^{n(H(Y|W)-R_K + 2\d_1(\e''))}+1$ in the definition of $E_1$, we have
\begin{align*}
H(Y^n|W^n, K) & \le H(Y^n, E_1, E_2|W^n, K) \\
& \le 2 + \P(E_1 = 0, E_2 = 0) H(Y^n |W^n, E_1 =0, E_2 =0, K) \\
& \quad + 2n\P(E_2 =1) \log|\Yc| + n\P(E_1 = 1)\log|\Yc| \\
& \le n(H(Y|W)- R_K+ \d(\e))
\end{align*}
for $n$ sufficiently large. The final step also uses the assumption that $R_K \le H(Y|W)$ and hence, $a = 1+ 2^{n(H(Y|W)-R_K + 2\d_1(\e''))} \le 2^{n(H(Y|W)-R_K + \d_2(\e''))}$ for $n$ sufficiently large.
\section{Proof of Proposition~\ref{propyz_out}} \label{appen:propyz_out}
Given a $(n, 2^{nR})$ code that achieves $(D + \e_n, \Delta+ \e_n)$, define the auxiliary random variables \\$V_i = (M, Y^{i-1},Z^{i-1},Y_{i+1}^n, Z_{i+1}^n)$ for $i \in [1:n]$. We have{\allowdisplaybreaks
\begin{align*}
nR & \ge H(M) \\
& \ge I(X^n; M|Y^n, Z^n) \\
& = \sum_{i=1}^n I(X_i;M|Y^n, Z^n, X^{i-1}) \\
& \stackrel{(a)}{=} \sum_{i=1}^n I(X_i;M, X^{i-1}, Y^{i-1},Z^{i-1},Y_{i+1}^n, Z_{i+1}^n|Y_i, Z_i) \\
&\ge \sum_{i=1}^n I(X_i; V_i|Y_i, Z_i),
\end{align*}}
where $(a)$ follows from the fact that the sources are i.i.d.. Next, for the information leakage rate{\allowdisplaybreaks
\begin{align*}
 n\Delta + n\e_n &= I(X^n; M, Z^n) \\
 & = I(X^n; Z^n) + I(X^n;M|Z^n) \\
& = I(X^n; Z^n) + I(X^n;M, Y^n|Z^n) - I(X^n; Y^n|M,Z^n) \\
& = I(X^n; Z^n) + I(X^n; Y^n|Z^n)+ I(X^n;M|Y^n, Z^n) - I(X^n; Y^n|M,Z^n) \\
& = I(X^n; Z^n) + I(X^n;M|Y^n, Z^n)+ I(X^n; Y^n|Z^n)- I(M, X^n; Y^n|Z^n) + I(M;Y^n|Z^n)  \\
& = I(X^n; Z^n) + I(X^n;M|Y^n, Z^n)- I(M; Y^n|Z^n, X^n) + I(M;Y^n|Z^n)\\
& \ge \sum_{i=1}^n (I(X_i;Z_i) + I(X_i;V_i|Y_i,Z_i) - H(Y_i|Z_i,X_i)).
\end{align*}}
Next, we let $Q\sim \U[1:n]$ and define $V = (V_Q, Q)$, $X_Q = X$, $Y_Q = Y$, $Z_Q = Z$. For the distortion, we have {\allowdisplaybreaks
\begin{align*}
D+ \e_n & \ge \frac{1}{n}\E \sum_{i=1}^nd(X_i, \xh_i(Z^n,Y^n, M)) \\
& = \frac{1}{n}\E \sum_{i=1}^nd(X_i, \xh_i(V_i, Z_i,Y_i)) \\
& = \E d(X, \xh(V,Z,Y)).
\end{align*}}
Then, noting that $\e_n \to 0$ as $n \to \infty$ and using the i.i.d. property of the source and S.I., we obtain the bounds stated in the proposition. The Markov Chain condition $V-(X,Y)-Z$ follows from the definition of $V$ and is easy to verify. The cardinality bound for $V$ follows from standard arguments~\cite[Appendix C]{El-Gamal--Kim2010} and we omit it here.
\section{Proof of Proposition~\ref{prop4}} \label{rhach}
We first state the following lemma that we will use in our analysis of information leakage rate in our proof. The proof of this lemma is given in Appendix~\ref{appenkey2}. 
\begin{lemma} \label{lemkey2}
Fix $\e>0$. Let $U^n(l), l \in [1:2^{n\Rt}]$ be generated according to $\prod_{i=1}^n p(u_i)$. Let $\Wt^n$ be a random variable and assume that there exists a random variable $L \in [1:2^{n\Rt}]$ such that $\P((U^n(L), \Wt^n) \in \aep) \to 1$ as $n \to \infty$. Bin the $U^n(l)$ sequences uniformly at random to $2^{nR_K}$ bins, $\Bc(k)$, $k \in [1:2^{nR_K}]$. Let $K$ be the index such that $U^n(L) \in \Bc(K)$. For $n$ sufficiently large, let $\d_1(\e')$ be a function of $\e'$, where $\e' >\e$, such that: $\d_1(\e') \to 0$ as $\e' \to 0$; and $\P((U^n(1), \wt^n)\in \aepvar) \ge 2^{-n(I(U;\Wt)+ \d_1(\e'))}$ for $\wt^n \in \aep(\Wt)$\footnote{The existence of $\d_1(\e')$ for $n$ sufficiently large follows from the conditional typical lemma~\cite[Chapter 2]{El-Gamal--Kim2010}}. Then, for $n$ sufficiently large and $\Rt - I(U;\Wt) -R_K >\d_1(\e')$, 
\begin{align*}
H(L|K,\Wt^n) \le n(\Rt - R_K - I(U;\Wt) + \d(\e)).
\end{align*}
\end{lemma}

We will also use Lemma~\ref{lem2}, stated in Appendix~\ref{appen:prop2}, in our analysis. 

Now, we turn to the achievability proof. We assume in our proof that $I(U_h;Y|Z) \ge I(U_h;Y|X)$. The proof when the inequality is reversed follows the same arguments, and is omitted. 
\subsection*{Codebook generation}
We start with the codebook generation at the helper.
\begin{itemize}
\item Generate $2^{n(I(U_h;Y) + 3\d(\e))}$ $U_h^n(l_h)$, $l_h \in [1:2^{n(I(U_h;Y) + 3\d(\e))}]$ sequences according to $\prod_{i=1}^n p(u_{h,i})$.
\item Partition the codewords to $2^{n(I(U_h;Y|Z) + 5\d(\e))}$ bins, $\Bc_h(m_h)$, $m_h \in [1:2^{n(I(U_h;Y|Z) + 5\d(\e))}]$. 
\end{itemize} 
Next, we turn to the codebook generation at the encoder
\begin{itemize}
\item Generate $2^{n(I(U;X,U_h) + \d(\e))}$ $U^n(l_0)$ sequences according to $\prod_{i=1}^n p(u_i)$, $l_0 \in [1: 2^{n(I(U;X,U_h) + \d(\e))}]$.
\item For each $u^n(l_0)$ sequence, generate $2^{n(I(V;X,U_h|U) + \d(\e))}$ $V^n(l_1, l_0)$ sequences according to \\$\prod_{i=1}^n p(v_i|u_i)$, $l_1 \in [1: 2^{n(I(V;X,U_h|U) + \d(\e))}]$.  
\item Partition the set of $U^n$ sequences to $2^{n(I(U;X|U_h, Z) + 2\d(\e))}$ bins, $\Bc_{\rm RD}(m_0)$, \\$m_0 \in [1: 2^{n(I(U;X|U_h,Z) + 2\d(\e))}]$.
\item For each $l_0$, partition the set of $V^n$ sequences to $2^{n(I(V;X|U_h,U, Z) + 2\d(\e))}$ bins, $\Bc_{\rm RD}(m_1,l_0)$, \\$m_1 \in [1: 2^{n(I(V;X|U_h,Z,U) + 2\d(\e))}]$.
\end{itemize}
We now turn to the key generation codebook, $\Cc_{\rm K}$, which has only a single step. We assume that $I(U_h;Y) - I(U_h; X,W,V,U) > 0$. Otherwise, the $U_h^n(l)$ codewords are not used to generate a secret key.
\begin{itemize}
\item Randomly and uniformly bin the set of $U_h^n(l)$ sequences to $2^{nR_K}$ bins, $\Bc_{\rm K}(m_k)$, $m_k \in [1:2^{nR_K}]$ and $R_K\le I(U_h;Y) - I(U_h; X,W,V,U) + \d(\e)$.
\end{itemize}
We use $\Cc:=\{\Cc_{\rm RD}, \Cc_{\rm K}\}$ to denote the combined codebook.
\subsection*{Encoding}
Encoding at the helper.
\begin{itemize}
\item Given sequence $y^n$, the helper looks for a codeword $u_h^n(l_h)$ such that $(u^n(l_h), y^n) \in \aep$. If there is more than one such codeword, it selects a codeword uniformly at random from the set of all jointly typical codewords. If there is none, it selects an index uniformly at random from the set of all possible indices.
\item Note that we have $\P((U^n_h(L_h), Y^n) \in \aep) \to 1$ as $n \to\infty$. Further, from the conditional typicality lemma~\cite[Chapter 2]{El-Gamal--Kim2010} and the Markov relation $U_h - Y- (X,Z,W)$, we have\\ $\P((U^n_h(L_h), Y^n, X^n, Z^n, W^n) \in \aepvar) \to 1$ as $n \to \infty$. 
\item The helper finds $m_h$ such that $u_h^n(l_h) \in \Bc(m_h)$.
\item Next, using its own independent randomness, the helper generates an additional key $m_k'$ uniformly distributed over the set $[1:2^{nR_K'}]$.
\item The helper sends out $m_h$ and $m_k'$, resulting in a rate that is less than or equal $R_h$. 
\end{itemize}
\subsection*{Decoding helper's message at the encoder}
The encoder first decodes the helper's message. That is, it looks for the unique $u^n_h(\lh_h)$ such that $(u^n(\lh_h^n), x^n) \in \aep$ and $u^n(\lh_h) \in \Bc(m_h)$. Following standard analysis and the rates given for $m_h$ and $l_h$, the probability of error in decoding $l_h$ goes to zero as $n \to \infty$ since $I(U_h;Y|X) \le I(U_h;Y|Z)$. 
\subsection*{Encoding at the encoder}
\begin{itemize}
\item Given sequences $(x^n, u_h^n(\lh_h))$, the encoder first looks for a sequence $u^n(l_0)$ such that \\$(u^n(l_0),x^n, u_h^n(\lh_h))\in \aep$. If there is more than one such sequence, the encoder selects one sequence uniformly at randomly from the set of jointly typical $u^n$ sequences. If there is none, the encoder randomly and uniformly selects a sequence $u^n$ from the set of all sequences.
\item Next, the encoder looks for a $v^n(l_1,l_0)$ such that $(v^n(l_1,l_0),u^n(l_0),x^n, u_h^n(\lh_h))\in \aep$. If there is more than one such sequence, the encoder selects one sequence uniformly at random from the set of jointly typical $v^n$ sequences. If there is none, the encoder randomly and uniformly selects a sequence $v^n$ from the set of all sequences.
\item The encoder then looks for the index $m_0$ and $m_1$ such that $u^n(l_0) \in \Bc(m_0)$ and $v^n(l_1,l_0) \in \Bc(m_1, l_0)$. 
\item Next, it splits the index $m_1$ into three parts, $m_{1s} \in [1:2^{nR_K}]$, $m_{1s}' \in [1:2^{nR_K'}]$ and \\$m_{1o} \in[1:2^{n(I(V;X|U,Y) + 2\d(\e) - R_K-R_K')}]$.
\item The encoder then looks for the index $m_k$ such that $y^n \in \Bc_{\rm K}(m_k)$.
\item Finally, the encoder sends out the indices $m_0$, $m_{1o}$, $m_{1s}\oplus m_{k}$ and $m_{1s}'\oplus m_{k}'$, resulting in a rate of $I(X;U,V|Y) + 4\d(\e)$. Note here that the constraints on $R_K$ and $R_K'$ guarantee the feasibility of the secret key scrambling operations ($m_{1s}\oplus m_{k}$ and $m_{1s}'\oplus m_{k}'$). 
\end{itemize}
\subsection*{Probability of error in encoding}
In our achievability scheme, we require that $\P((U_h^n(\Lh_h), U^n(L_0), V^n(L_0, L_1), X^n, Y^n, Z^n, W^n) \in \aep) \to 1$ as $n \to \infty$. Let $\Ec_1$ denote the event $(U_h^n(\Lh_h), U^n(L_0), V^n(L_0, L_1), X^n, Y^n, Z^n, W^n) \notin \aep$. Therefore, $\P(\Ec_1)$ denotes the probability of overall encoder error. Let $\Ec_0$ denote the event that $\{(U_h^n(L_h), U^n(L_0), X^n, Y^n, Z^n, W^n) \notin \aepvar\} \cup \{\hat{L}_h \neq L_h\}$. We know from the preceding analysis that $\P(\Ec_0) \to 0$ as $n \to \infty$ since $\P(\Lh_h \neq L_h) \to 0$ and $\P((U^n_h(L_h), Y^n, X^n, Z^n, W^n) \in \aepvar) \to 1$ as $n \to \infty$. To show that $\P(\Ec_1) \to 0$ as $n \to \infty$, it remains to show that $\P(\Ec_0^c \cap \Ec_1) \to 0$ as $n \to \infty$. To do so, we will use the Markov lemma in~\cite[Chapter 12]{El-Gamal--Kim2010} stated as follow.
\begin{lemma}[Markov Lemma]
Suppose $\Xt \to \Yt \to \Zt $. Let $(\xt^n, \yt^n) \in \aepvar$ and $\Zt^n \sim p(\zt^n|\yt^n)$, where the conditional pmf $p(\zt^n|\yt^n)$ satisfies the following conditions
\begin{enumerate}
\item $\P((\yt^n, \Zt^n) \in \aepvar) \to 1 $ as $n \to \infty$;
\item for every $\zt^n \in \aepvar(\Zt|\yt^n)$ and $n$ sufficiently large
\begin{align*}
2^{-n(H(\Zt|\Yt) + \d(\e')) }\le p(\zt^n|\yt^n) \le 2^{-n(H(\Zt|\Yt) - \d(\e')) }.  
\end{align*}
\end{enumerate}
Then, if $\e'$ is sufficiently small compared to $\e$, $\P((\xt^n, \yt^n, \Zt^n) \notin \aep) \to 0$ as $n \to \infty$.
\end{lemma}

Next, let $(\Xt^n, \Yt^n) = (U^n_h(L_h), Y^n, X^n, Z^n, W^n)$, $\Yt^n = (U^n_h(L_h), X^n)$ and \\$\Zt^ n = (V^n(L_0,L_1), U^n(L_0))$, we have
\begin{align*}
\P(\Ec_1 \cap \Ec_0^c) &\le \P(\Ec_1|\Ec_0^c) \\
&= \sum_{(\yt^n, \xt^n) \in \aepvar} \P((\xt^n, \yt^n)|\Ec_0^n)\P(\Ec_1|(\xt^n, \yt^n)).
\end{align*} 
Consider now the term $\P(\Ec_1|(\xt^n, \yt^n)) = \P((V^n(L_0,L_1), U^n(L_0), \xt^n, \yt^n) \notin \aep)$. Observe from the encoding process that $(V^n(L_0,L_1), U^n(L_0)) \to \Yt^n \to \Xt^n$. Hence, we now apply the Markov lemma to show that $\P((V^n(L_0,L_1), U^n(L_0), \xt^n, \yt^n) \notin \aep) \to 0$ for every $(\xt^n, \yt^n)\in \aepvar$. Condition 1 of the Markov lemma holds since from the rates given, codebook generation process, encoding process and standard analysis using the covering lemma of~\cite[Chapter 3]{El-Gamal--Kim2010}, $\P(((V^n(L_0,L_1), U^n(L_0), \yt^n) \in \aepvar) \to 1$ as $n \to \infty$. Next, we check that the second condition holds. The analysis closely follows that used in~\cite[Chapter 12, Lemma 12.3]{El-Gamal--Kim2010}, and we omit the details here.
\begin{align*}
\P(V^n(L_0,L_1) = v^n, U^n(L_0, L_1)=u^n|\yt^n) &= \P(U^n(L_0)=u^n|u^n_h, x^n) \P(V^n(L_0,L_1)=v^n|u^n, u^n_h, x^n) \\
& \stackrel{\stackrel{(a)}{.}}{=}2^{-nH(U|X,U_h)}\P(V^n(L_0,L_1)=v^n|u^n, u^n_h, x^n) \\
& \stackrel{\stackrel{(b)}{.}}{=}2^{-nH(U|X,U_h)}2^{-nH(V|U,X,U_h)} \\
& \stackrel{.}{=}2^{-nH(U,V|X,U_h)}.
\end{align*}
$(a)$ follows the same analysis as in~\cite[Chapter 12, Lemma 12.3]{El-Gamal--Kim2010}. $(b)$ also follows from an analysis similar to that in~\cite[Chapter 12, Lemma 12.3]{El-Gamal--Kim2010}, but conditioned on $U^n(L_0) = u^n$. 

Hence, $\P(\Ec_1|(\xt^n, \yt^n)) \to 0$ as $n \to \infty$ and therefore, $\P(\Ec_1|\Ec_0^c) \to 0$ as $n \to \infty$. We note here that our analysis also implies that $\P((U_h^n(L_h), U^n(L_0), V^n(L_0, L_1), X^n, Y^n, Z^n, W^n) \in \aep) \to 1$ as $n \to \infty$. This fact will be used in our analysis of information leakage rate. 
 
\subsection*{Decoding and analysis of distortion}
\begin{itemize}
\item The decoder first decodes the codeword from the helper by looking for an unique $u_h^n(\lh_h)$ such that $(u_h^n(\lh_h), z^n) \in \aep$, and $u_h^n(\lh_h) \in \Bc_h(m_h)$. The probability of error in this step goes to zero with $n$ since $R_h > I(U_h;Y|Z)$. 
\item The decoder next looks for the $\mh_k$ such that $u^n_h(\lh_h) \in \Bc_{\rm K}(\mh_k)$. 
\item It then unscrambles the indices $m_{1s}$ and $m_{1s}'$ by unscrambling $m_{1s}\oplus m_k$ and $m_{1s}'\oplus m_k'$ using $\mh_k$ and $m_k'$ respectively.
\item Finally, the decoder decodes the codewords $u^n(L_0)$ and $v^n(L_0,L_1)$ using the indices $m_0$ and $m_1$ by successive decoding (see decoding and analysis of probability of error in proof of Proposition~\ref{prop2}).
\end{itemize}
The analysis of the probability of error follows quite straightforwardly from the analysis for a similar setting in~\cite{Permuter--Steinberg--Weissman2010}, and we will omit it here. Finally, for the distortion constraint, similar to the proof in Proposition~\ref{prop2}, we note that since the probability of encoding error or decoding error goes to zero as $n \to \infty$, the expected distortion, averaged over codebooks, is less than or equal to $D + \d(\e)$ as $n \to \infty$~\cite[Chapter 3]{El-Gamal--Kim2010}. 

\subsection*{Analysis of information leakage rate}
For notational convenience, we will use $\d(\e)$ to denote all terms that go to zero as $\e \to 0$, or $n\to \infty$. We will also suppress the indices for the codewords. Hence, $U^n_h(L_h)=U^n_h$, $U^n(L_0)=U^n$ and $V^n(L_0, L_1) = V^n$. Note also that in our analysis, the manipulation of mutual information and entropy quantities will use the three Markov relations: MC1: $U_h-Y-(X,Z,W)$, MC2: $(V,U)- (X,U_h) - (Y,Z,W)$ and MC3: $(V,U,U_h) - (X,Y) - (W,Z)$ ) stated in the Proposition. For brevity, we will not state these relations explicitly in the analysis, but indicate by the labels (MC1, MC2, MC3) whether MC1, MC2 or MC3 is used in the steps in the analysis.{\allowdisplaybreaks
\begin{align}
n\Delta & = I(X^n; W^n, M_0, M_{1o}, M_{1s}\oplus M_K, M_{1s}'\oplus M_K'| \Cc) \nonumber\\
& \le I(X^n; W^n, L_0, M_{1o}, M_{1s}\oplus M_K, M_{1s}'\oplus M_K'| \Cc) \nonumber\\
& = I(X^n, Y^n; W^n, L_0, M_{1o}, M_{1s}\oplus M_K, M_{1s}'\oplus M_K'|\Cc) \nonumber\\
& \qquad- I(Y^n; W^n, L_0, M_{1o}, M_{1s}\oplus M_K, M_{1s}'\oplus M_K'|X^n, \Cc). \label{ilb1}
\end{align}}
Similar to Proposition~\ref{prop2}, we analyze the two terms in~\eqref{ilb1} separately. For the first term, an additional term comes up due to independent randomness. {\allowdisplaybreaks
\begin{align}
& I(X^n, Y^n; W^n, L_0, M_{1o}, M_{1s}\oplus M_K, M_{1s}'\oplus M_K'|\Cc) \nonumber\\
& = H(W^n, L_0|\Cc) + H(M_{1o}, M_{1s} \oplus M_K, M_{1s}'\oplus M_K'|L_0, W^n, \Cc) \nonumber\\
& \quad -  H(W^n, L_0, M_{1o}, M_{1s}\oplus M_K, M_{1s}'\oplus M_K'|X^n, Y^n, \Cc) \nonumber\\
& \stackrel{(a)}{\le} H(W^n, L_0|\Cc) + H(M_{1o}, M_{1s} \oplus M_K, M_{1s}'\oplus M_K'|L_0,\Cc)\nonumber \\
& \quad -  H(W^n|X^n, Y^n, \Cc)- H(M_{1s}'\oplus M_K') \nonumber\\
& \le H(L_0|\Cc) + H(W^n|L_0, \Cc) + n I(V;X|U,Z, U_h)  -  H(W^n|X^n, Y^n, \Cc) -  nR_K'  + n \d(\e)\nonumber\\
& \le H(L_0|\Cc) + H(W^n|U^n(L_0)) + n I(V;X|U,Z, U_h)  -  H(W^n|X^n, Y^n, \Cc) -  nR_K'  + n \d(\e)\nonumber\\
& \stackrel{(b)}{\le} H(L_0|\Cc) + nH(W|U) + n I(V;X|U,Z, U_h)  -  H(W^n|X^n, Y^n, \Cc) -  nR_K'  + n \d(\e)\nonumber\\
&\le nI(U;X, U_h) + nH(W|U)  + n I(V;X|U,Z, U_h) -  nH(W|X, Y)-  nR_K' + n \d(\e)\nonumber\\
& \stackrel{MC1}{=} nI(U;X, U_h, Y)  + nH(W|U)  + n I(V;X|U,Z, U_h) -  nH(W|X, Y)-  nR_K' + n \d(\e)\nonumber\\
& \stackrel{MC3}{=} nI(U;X, Y) + nI(U; U_h|X,Y)  + nH(W|U) + n I(V;X|U,Z, U_h) \nonumber\\
& \qquad -  nH(W|X, Y,U)-  nR_K' + n \d(\e)\nonumber\\
& = nI(W, U; X, Y)  + nI(U; U_h|X,Y)+ n I(V;X|U,Z, U_h) - nR_K' + n \d(\e)\nonumber\\
& = nI(W, U; X) + nI(W, U; Y|X) + nI(U; U_h|X,Y)+ n I(V;X|U,Z, U_h) - nR_K'+ n \d(\e). \label{ilb2}
\end{align}}
$(a)$ uses the fact that $M_{1s}'\oplus M_K'$ is independent of all other random variables due to $M_K'$ being uniformly distributed and independent of other random variables. $(b)$ follows from application of Lemma~\ref{lem2} (see proof of Proposition~\ref{prop2} in Appendix~\ref{appen:prop2}) to the third term. The conditions required for application of Lemma~\ref{lem2} are satisfied as, from the rates given and the encoding process, $\P((U_h^n, U^n, V^n, X^n, Y^n, Z^n, W^n) \in \aep) \to 1$ as $n \to \infty$. 

For the second term, we have{\allowdisplaybreaks
\begin{align}
& - I(Y^n; W^n, L_0, M_{1o}, M_{1s}\oplus M_K,  M_{1s}'\oplus M_K'|X^n, \Cc)\nonumber\\
& = - H(Y^n|X^n) + H(Y^n|X^n, W^n, L_0, M_{1o}, M_{1s}\oplus M_K, \Cc) \nonumber\\
& \le -nH(Y|X) + H(Y^n, L_h, L_1|X^n, W^n, L_0, M_{1o}, M_{1s}\oplus M_K, \Cc) \nonumber\\ 
& = -nH(Y|X) + H(L_1|X^n, W^n, L_0, M_{1o}, M_{1s}\oplus M_K, \Cc)\nonumber\\
& \quad + H(Y^n|X^n, W^n, L_0, L_1, M_k,L_h, \Cc)+H(L_h|X^n, W^n, L_0, L_1, M_K, \Cc) \nonumber \\
& \le -nH(Y|X) + H(L_1|X^n, W^n, L_0, M_{1o}, M_{1s}\oplus M_K, \Cc)\nonumber\\
& \quad + H(Y^n|X^n, W^n, V^n, U^n, U_h^n)+H(L_h|X^n, W^n, V^n, U^n, M_K)  \nonumber\\
& \stackrel{(a)}{\le}  -nH(Y|X) + H(L_1|X^n, W^n, L_0, M_{1o}, M_{1s}\oplus M_K, \Cc)\nonumber\\
& \quad + n H(Y|U,V,X,W, U_h)+H(L_h|X^n, W^n, V^n, U^n, M_K) + n\d(\e) \nonumber\\
& \stackrel{(b)}{\le}  -nH(Y|X) + H(L_1|X^n, W^n, L_0, M_{1o}, M_{1s}\oplus M_K, \Cc)\nonumber\\
& \quad + n H(Y|U,V,X,W, U_h)+ nI(U_h;Y)-nI(U_h; X,W, U,V)- nR_K + n\d(\e) \nonumber\\
& \stackrel{(c)}{\le} -nH(Y|X) + nI(V;Y|U, X,W) + nI(V,U, X; U_h|Y)\nonumber\\
& \quad + n H(Y|U,V,X,W, U_h)+ nI(U_h;Y)-nI(U_h; X,W, U,V)- nR_K + n\d(\e)\nonumber\\
& = -nH(Y|X) + nI(V;Y|U,X,W)+ nI(V,U; U_h|Y,X) \nonumber\\
& \quad + n H(Y|U,V,X,W, U_h)+ nI(U_h;Y)-nI(U_h; X,W, U,V)- nR_K + n\d(\e) \nonumber\\
& \le -nH(Y|X) + nI(V;Y|U,X,W)  + nI(V,U; U_h|Y,X) \nonumber\\
& \quad + n H(Y|U,V,X,W, U_h)+ nI(U_h;Y|X,W, U,V)- nR_K + n\d(\e)\nonumber\\
& = -nI(Y;U,W|X) + nI(V,U; U_h|Y,X)-nR_K +n \d(\e). \label{ilb3} 
\end{align}}
$(a)$ follows from application of Lemma~\ref{lem2} to the third term. It is again straightforward to verify that the conditions required for application of Lemma~\ref{lem2} are satisfied from the rates given and the encoding process. 
 
$(b)$ follows applying Lemma~\ref{lemkey2} to the last term, with $\Rt = I(U_h;Y) + 3\d(\e)$, $R_K \le I(U_h;Y) - I(U_h; X,W,U,V)$ and $\Wt = (X,W,U,V)$. The conditions required for application of Lemma~\ref{lemkey2} in $(b)$ follow from the rates given and the encoding process. 
 
 In $(c)$, we upper bound $H(L_1|X^n, W^n, L_0, M_{1o}, M_{1s}\oplus M_K, \Cc)$ as follow.{\allowdisplaybreaks
\begin{align*}
& H(L_1|X^n, W^n, L_0, M_{1o}, M_{1s}\oplus M_K, \Cc) \\
&\le H(L_1|X^n, W^n, L_0, \Cc) \\
&= H(L_1| L_0, \Cc) + H(X^n, W^n|L_0, L_1, \Cc) - H(X^n, W^n|L_0, \Cc) \\
& \le H(L_1| L_0, \Cc) + H(X^n, W^n|U^n, V^n) - H(X^n, W^n, L_0, \Cc) + H(L_0| \Cc) \\
& \stackrel{(i)}{\le} H(L_1, L_0|\Cc) + nH(X, W|U,V) - nH(X,W) - H(L_0|X^n, W^n, \Cc) +  n\d(\e) \\
& \le H(L_1, L_0|\Cc) + nH(X, W|U,V) - nH(X,W)- I(L_0;Y^n|X^n, W^n, \Cc)  + n\d(\e)\\
& \le H(L_1, L_0|\Cc) + nH(X, W|U,V) - nH(X,W)- nH(Y|X,W) + H(Y^n|U^n, X^n, W^n)  + n\d(\e) \\
& \stackrel{(ii)}{\le} H(L_1, L_0|\Cc) + nH(X, W|U,V) - nH(X,W)- nH(Y|X,W) + nH(Y|U, X, W) +  n\d(\e) \\
& \le nI(V,U;X, U_h) + nH(X, W|U,V) - nH(X,W)- nH(Y|X,W) + nH(Y|U, X, W) +  n\d(\e) \\ 
& = nI(V,U;X, U_h) - nI(U,V;X,W)- nI(U;Y|X,W) + n\d(\e)\\
& \stackrel{MC2}{=} nI(V,U;X, U_h, Y,W) - nI(U,V;X,W)- nI(U;Y|X,W) + n\d(\e)\\
& = nI(V,U; U_h, Y|X,W) - nI(U;Y|X,W) + n\d(\e)\\
& = nI(V;Y|U, X,W) + nI(V,U; U_h|Y,X) + n\d(\e)\\
& \stackrel{MC1}{=} nI(V;Y|U, X,W) + nI(V,U, X; U_h|Y)+ n\d(\e).
\end{align*}}
$(i)$ and $(ii)$ follow from application of Lemma~\ref{lem2}.

Combining the bounds for the two terms in~\eqref{ilb2} and~\eqref{ilb3} into~\eqref{ilb1} then leads to the upper bound on the information leakage rate, which then completes the proof of achievability for Proposition~\ref{prop4}.
\section{Proof of Lemma~\ref{lemkey2}} \label{appenkey2}
Define $N(\wt^n, k):= |\{l: U^n(l) \in \Bc(k), (U^n(l),\wt^n) \in \mathcal{T}_{\e'}^{(n)}\}|$. Define $E_1 = 1$ if $N(\Wt^n, K) > a$ and $0$ otherwise. Let $E_2 = 1$ if $(\Wt^n,U^n(L)) \notin \aep$ and $0$ otherwise. Observe that by assumption, $\P(E_2 =1) \to 0$ as $n \to \infty$. We now focus on upper bounding $E_1$.{\allowdisplaybreaks
\begin{align}
\P(E_1 =1) &\le \sum_{\wt^n \in \aep, k}\P(E_1=1, \Wt^n = \wt^n, K=k) + \P(\Wt^n \notin \aep) \nonumber \\
& \le \sum_{\wt^n \in \aep, k}\P(N(\wt^n, k)> a, \Wt^n =\wt^n, K= k) + \e_n \nonumber\\
& \le \sum_{\wt^n \in \aep, k}\P(N(\wt^n, k)> a) + \e_n. \label{eqp} 
\end{align}}
Now, we use a version of the Chernoff bound, found in~\cite[Appendix B]{El-Gamal--Kim2010}. Let $X_1, X_2, X_3, \ldots, X_m$ be i.i.d. binary random variables with $\P(X_j =1) = p$. Then,
\begin{align*}
\P\left(\sum_{j=1}^mX_j \ge m(1+ \d)p\right) \le \exp(-\d^2 mp /4)
\end{align*}
for $\d\in (0,1)$.
Now, let $m = 2^{n\Rt}$ and let $X_j$ be the indicator function of the event \\$\{U^n(j) \in \Bc(k), (U^n(j), \wt^n)\in \aepvar\}$. We note that $X_j$s are i.i.d. binary random variables since the binning is done uniformly at random and $U^n(j)$ is generated according to $\prod_{i=1}^n p(u_i)$ for all $j$. Next, since the binning is done uniformly at random, independent of all other random variables, $\P(X_j) = \P(U^n(j) \in \Bc(k))\mathbf{.}\P((U^n(j), \wt^n)\in \aepvar)$. Hence, 
\begin{align*}
p &= 2^{-nR_K}\P((U^n(j), \wt^n)\in \aepvar)\\
& \ge 2^{-nR_K}2^{-n(I(U;\Wt)+ \d_1(\e'))}
\end{align*}
for $n$ sufficiently large. The second step follows from the statement of lemma~\ref{lemkey2}, which, in turn, follows from the conditional typical lemma~\cite[Chapter 2]{El-Gamal--Kim2010}.

Applying the Chernoff bound to~\eqref{eqp} with $a = (1+ \d)mp$, we obtain
\begin{align*}
\P(E_1 = 1) &\le \sum_{\wt^n \in \aep, k} \exp(-\d^2 2^{n(\Rt-R_K- I(IU;\Wt) -\d_1(\e'))}  /4) \\
& \le |\aep(\Wt)|2^{nR_K}\exp(-\d^22^{n(\Rt-R_K- I(IU;\Wt) +\d_1(\e'))}  /4).
\end{align*}
By assumption, $\Rt-R_K- I(U;\Wt)>\d_1(\e')$ and hence, $\P(E_1 = 1) \to 0$ as $n \to \infty$.
We therefore have{\allowdisplaybreaks
\begin{align*}
H(L|K,\Wt^n) & \le H(L, E_1, E_2|\Wt^n, K) \\
& \le 2 + \P(E_1 = 0, E_2 = 0) H(L |\Wt^n, E_1 =0, E_2 =0, K) \\
& \quad + 2n\Rt\P(E_2 =1) + n\Rt\P(E_1 = 1) \\
& \le n(\Rt - R_K - I(U;\Wt) + \d(\e))
\end{align*}}
for $n$ sufficiently large.
\section{Proof of Proposition~\ref{coro6}} \label{appen:c}

For the converse, consider an $(n, 2^{nR}, 2^{nR_h})$ code achieving $(D + \e_n, \Delta+ \e_n)$. The lower bound on $R_h$ is trivial. For $R$, we have{\allowdisplaybreaks
\begin{align*}
nR &\ge H(M) \\
& \ge I(X^n;M|Z^n, M_h) \\
& = I(X^n;M, M_h|Z^n) - I(X^n; M_h|Z^n) \\
& \stackrel{(a)}{=} I(X^n;M, M_h|Z^n) \\
& = I(X^n;M, M_h, \Xh^n|Z^n) \\
& \ge \sum_{i=1}^n I(X_i; \Xh^n |Z^n, X^{i-1}) \\
& \stackrel{(b)}{=} \sum_{i=1}^n I(X_i; \Xh^n, Z_{i+1}^n, Z^{i-1}, X^{i-1}|Z_i) \\
& \ge \sum_{i=1}^n I(X_i; \Xh_i|Z_i).
\end{align*} }
In $(a)$, we used the Markov Chain assumption $Y-W-Z-X$. $(b)$ follows from the fact that sources are i.i.d.. 

For the information leakage rate, the lower bound $n\Delta + n\e_n\ge \sum_{i=1}^n I(X_i;W_i)$ is straightforward to show. We also have {\allowdisplaybreaks
\begin{align*}
n\Delta + nR_h + n\e_n& \ge I(X^n;M,W^n) + H(M_h) \\
& = I(X^n;W^n) + I(X^n; M|W^n) + H(M_h) \\
& \stackrel{(a)}{=} I(X^n;W^n) + I(Z^n, X^n; M|W^n) + H(M_h) \\
& \ge I(X^n;W^n) + I( X^n; M|Z^n, W^n) + H(M_h) \\
& \ge I(X^n;W^n) + I( X^n; M, W^n|Z^n) - I(X^n;W^n|Z^n) + H(M_h) \\
& \stackrel{(b)}{\ge} I(X^n;W^n) + I( X^n; M|Z^n)  + H(M_h) \\
& = I(X^n;W^n) + I(X^n; M, M_h|Z^n) - I(X^n; M_h|M, Z^n)+ H(M_h) \\
& \ge I(X^n;W^n) + I(X^n; M, M_h|Z^n) \\
& \ge I(X^n;W^n) +  \sum_{i=1}^n I(X_i; \Xh_i|Z_i). 
\end{align*}}
$(a)$ and $(b)$ follow from the Markov Chain assumption $Y-W-Z-X$. The last step follows the same arguments used in lower bounding $R$. Now, let $Q\sim \U[1:n]$ independent of other random variables and define $(X_Q,Y_Q,Z_Q, W_Q) = (X,Y,Z, W)$ and $\Xh_Q = \Xh$. We have
\begin{align*}
nR & \ge nI(X_Q; \Xh_Q|Z_Q, Q)\\
& \ge n I(X;\Xh|Z) \\
& \ge nR_{\rm SI-Enc}(D + \e_n).
\end{align*}
The last step follows from $\E \sum_{i=1}^n d(x_i, \xh_i)/n = \E d(X, \Xh) \le D + \e_n$ and the fact that $R_{\rm SI-Enc}(D + \e_n) = \min I(X;\Xh|Z)$, where we minimize over $p(\xh|x,z)$ satisfying $\E d(X, \Xh) \le D + \e_n$. Similarly, we have
\begin{align*}
n\Delta + nR_h + n\e_n& \ge n I(X;W) + nR_{\rm SI-Enc}(D + \e_n).
\end{align*}
Finally, noting that $\e_n \to 0$ as $n \to \infty$ and using the fact that $R_{\rm SI-Enc}(D)$ is continuous in $D$~\cite[Chapter 11]{El-Gamal--Kim2010}, we obtain the stated bound in the Proposition. This completes the proof of converse.

For the achievability, we use Proposition~\ref{prop4} and set $U_h = \emptyset$ and $U = \emptyset$. Using the assumption that $\Rc_{\rm SI-Enc}(Z) = \Rc_{\rm WZ}(Z)$, there exists an auxiliary random variable $V^*$ such that $V^* - X- Z$, $I(V^*;X|Z) = R_{\rm SI-Enc}(D)$ and $\E d(X, \xh(V^*, Z)) \le D$ for some reconstruction function $\xh(V^*, Z)$. We set $V =V^*$ in Proposition~\ref{prop4}. It is now straightforward to verify that Proposition~\ref{prop4} achieves the stated R.D.I. region. 
\section{Proof of Proposition~\ref{qg2}} \label{appen:d}
For the converse, using the fact that $Y-W-Z-X$ and following the same steps as the proof of converse for Proposition~\ref{coro6} in Appendix~\ref{appen:c}, we can show that 
\begin{align*}
R_h &\ge 0,\\
R &\ge I(X;\Xh|Z), \\
\Delta &\ge \max\{I(X;W), I(X;W) + I(X;\Xh|Z) - R_h\},
\end{align*}
for $\P_{\Xh|X,Z}$ satisfying $\E (X -\Xh)^2 \le D$ constitute an outer bound to the R.D.I. region. Now, using the condition that $\E (X - \Xh)^2 \le D$, we have
\begin{align*}
I(X;\Xh|Z) &\ge h(X|Z) - h(X -\Xh) \\
& \ge \frac{1}{2}\log\left(\frac{\sigma_X^2\sigma_A^2}{(\sigma_X^2 + \sigma_A^2)D}\right).
\end{align*}
Hence, the outer bound reduces to
\begin{align*}
R_h &\ge 0, \\
R & \ge [\frac{1}{2}\log\left(\frac{\sigma_X^2\sigma_A^2}{(\sigma_X^2 + \sigma_A^2)D}\right)]^+, \\
\Delta &\ge \max\{\frac{1}{2}\log\left(\frac{\sigma_X^2+\sigma_A^2+ \sigma_B^2}{\sigma_A^2+ \sigma_B^2}\right), \frac{1}{2}\log\left(\frac{\sigma_X^2+\sigma_A^2+ \sigma_B^2}{\sigma_A^2+ \sigma_B^2}\right) + \frac{1}{2}\log\left(\frac{\sigma_X^2\sigma_A^2}{(\sigma_X^2 + \sigma_A^2)D}\right) - R_h\}.
\end{align*}

For the achievability, using Proposition~\ref{prop4}, we set $U = U_h = \emptyset$ and let $\sigma_{X|Z}^2 = \frac{\sigma_X^2\sigma_A^2}{(\sigma_X^2 + \sigma_A^2)}$. We then set $V = X + V'$, where $V' \sim N(0, \frac{\sigma_{X|Z}^2}{\sigma_{X|Z}^2 -D})$ for $D \le \sigma_{X|Z}^2$. Then, we have $V-X-(Z,W,Y)$ and it is straightforward to verify that the Proposition~\ref{prop4} achieves R.D.I. region with this choice of auxiliary random variables. The case of $D > \sigma_{X|Z}^2$ is straightforward and this completes the proof.   
\section{Proofs of converse for Corollaries~\ref{coro2},~\ref{coro5} and~\ref{coro8} under block log-loss constraint} \label{appen:amp}
\subsection*{Proof of converse for Corollary~\ref{coro2} under block log-loss}
Given a $(n, 2^{nR})$ code that achieves $(D+ \e_n, \Delta+ \e_n)$, it is easy to show using inequality~\eqref{ent} that{\allowdisplaybreaks
\begin{align*}
nR &\ge nH(X|Y) - n(D+\e_n), \\
n\Delta + n\e_n &\ge n I(X;Z).
\end{align*}}
Further, we have{\allowdisplaybreaks
\begin{align*}
n\Delta + n\e_n & = I(X^n; Z^n) + I(X^n;M| Z^n) \\
& = I(X^n; Z^n) + I(X^n, Y^n;M|Z^n) - I(Y^n; M|X^n, Z^n) \\
& \ge nI(X; Z) + I(X^n;M|Y^n, Z^n) - nH(Y|X, Z) \\
& \ge nI(X; Z) + nH(X|Y) - nD -n\e_n - nH(Y|X, Z).
\end{align*}}
The last step uses the Markov Chain assumption $X-Y-Z$ and inequality~\eqref{ent} on $H(X^n|Y^n, M)$. Noting that $\e_n \to 0$ as $n \to \infty$ then completes the proof of converse.
\subsection*{Proof of converse for Corollary~\ref{coro5} under block log-loss}
Given a $(n, 2^{nR})$ code that achieves $(D+ \e_n, \Delta+ \e_n)$, we have, using inequality~\eqref{ent}
\begin{align*}
nR &\ge nH(X|Y,Z) - n(D+\e_n), \\
n\Delta + n\e_n &\ge n I(X;Z).
\end{align*}
Further, following the same arguments to the proof of converse for Corollary~\ref{coro2} under block log-loss in the previous section,
\begin{align*}
n\Delta + n\e_n & \ge nI(X; Z) + nH(X|Y,Z) - nD -n\e_n - nH(Y|X, Z).
\end{align*}
Noting that $\e_n \to 0$ as $n \to \infty$ then completes the proof of converse.

\subsection*{Proof of converse for Corollary~\ref{coro8} under block log-loss}
Given a $(n, 2^{nR}, 2^{nR_h})$ code that achieves $(D+ \e_n, \Delta+ \e_n)$, we have{\allowdisplaybreaks
\begin{align*}
nR_h \ge 0,
n\Delta + n\e_n &\ge n I(X;Z).
\end{align*}}
Further, {\allowdisplaybreaks
\begin{align*}
n R &\ge I(X^n;M|M_h, Z^n) \\
& \ge H(X^n|Z^n, M_h) - H(X^n|Z^n, M_h, M) \\
& \ge nH(X|Z) - nD - n\e_n.
\end{align*}}
The last step follows from $Y-W-Z-X$ and inequality~\eqref{ent}. For the information leakage rate, following the proof of converse for Proposition~\ref{coro6} in Appendix~\ref{appen:c} we have
\begin{align*}
n\Delta + nR_h + n\e_n & \ge I(X^n;W^n) + I(X^n; M, M_h|Z^n) \\
& \ge nI(X;W) + nH(X|Z) - nD - n\e_n.
\end{align*}
The last step follows from inequality~\eqref{ent}. Noting that $\e_n \to 0$ as $n \to \infty$ then completes the proof of converse.
\end{document}